\documentclass[iop]{emulateapj}

\shorttitle{Radio Emission from QSOs}
\shortauthors{Condon et al.}

\begin{document}

\title{AGN and Starburst Radio Emission from Optically Selected QSOs}

\author{J.~J.~Condon\altaffilmark{1}, K.~I.~Kellermann\altaffilmark{1},
and Amy E.~Kimball\altaffilmark{1}}
\affil{National Radio Asronomy Observatory, 520 Edgemont Road, 
Charlottesville, VA 22903, USA}

\author{\v{Z}eljko Ivezi\'{c}}
\affil{Department of Astronomy, University of Washington, Box 351580,
  Seattle, WA 98195, USA}

\author{R.~A.~Perley\altaffilmark{1}}
\affil{National Radio Astronomy Observatory, Socorro, NM 87801, USA}
\email{jcondon@nrao.edu}

\altaffiltext{1}{The National Radio Astronomy Observatory is a
  facility of the National Science Foundation operated under
  cooperative agreement by Associated Universities, Inc.}

\begin{abstract}
We used the 1.4 GHz NVSS to study radio sources in two color-selected
QSO samples: a volume-limited sample of 1313 QSOs defined by $M_{\rm
  i} < -23$ in the redshift range $0.2 < z < 0.45$ and a
magnitude-limited sample of 2471 QSOs with $m_{\rm r} \leq 18.5$ and
$1.8 < z < 2.5$.  About 10\%  were detected above the 2.4
mJy NVSS catalog limit and are powered primarily by AGNs.  The space
density of the low-redshift QSOs evolves as $\rho \propto (1+z)^6$.
In both redshift ranges the flux-density distributions and luminosity
functions of QSOs stronger than 2.4 mJy are power laws, with no
features to suggest more than one kind of radio source.  Extrapolating
the power laws to lower luminosities predicts the remaining QSOs
should be extremely radio quiet, but they are not. Most were detected
statistically on the NVSS images with median peak flux densities
$S_{\rm p}({\rm mJy~beam}^{-1}) \approx 0.3$ and $0.05$ in the low-
and high-redshift samples, corresponding to spectral luminosities
$\log[L_{\rm 1.4\,GHz}({\rm W~Hz}^{-1})] \approx 22.7$ and 24.1,
respectively.  We suggest that the faint radio sources are powered by
star formation at rates $\dot{M} \sim 20 M_\sun {\rm
  ~yr}^{-1}$ in the moderate luminosity (median $\langle M_{\rm i}
\rangle \approx -23.4$) low-redshift QSOs and $\dot{M} \sim 500 M_\sun {\rm
  ~yr}^{-1}$ in the very luminous ($\langle M_{\rm i} \rangle \approx
-27.5$) high-redshift QSOs.  Such luminous starbursts [$\langle
  \log(L_{\rm IR}/L_\sun) \rangle \sim 11.2$ and 12.6, respectively]
are consistent with ``quasar mode'' accretion in which cold gas flows
fuel both AGN and starburst.
\end{abstract}


\section{Introduction}\label{introsec}

The radio emission from quasi-stellar objects (QSOs) has puzzled
astronomers for nearly 50 years.  The first QSOs to be recognized
contain very strong radio sources similar to those powered by the
active galactic nuclei (AGNs) of the most luminous radio galaxies, but
$\sim90$\% of bright optically selected QSOs are fainter than 1~mJy at
1.4~GHz.  Also, the flux-density distribution of radio-loud QSOs is
quite different from that of radio galaxies, and might even be
bimodal.  These differences led to longstanding debates about whether
QSOs experience the same cosmological evolution as radio galaxies and
whether there are two distinct types of radio sources in QSOs.  In
this paper, we use direct detections of radio sources stronger than
2.4~mJy and statistical detections of fainter sources in two large
samples of QSOs to argue that (1) radio-loud QSOs are similar to
powerful radio galaxies and do evolve, despite their differing
flux-density distributions, and (2) radio-quiet QSOs really are
different from radio-loud QSOs because most of their radio emission is
powered by starbursts instead of AGNs.  Since the problem of radio
emission from QSOs has such a long and complicated history, we begin
with a short review to set the stage and define common terms.

\subsection{Quasars, QSGs, and QSOs}

\citet{haz63}, \citet{sch63}, and \citet{mat63} found the first
quasi-stellar radio sources (quasars or QSSs) by optically identifying
redshifted ``stars'' with radio sources whose luminosities are ``not
markedly different from other known strong radio sources like Cygnus A
or 3C 295'' \citep{gre63} in elliptical galaxies.  The first optically
selected quasi-stellar galaxies (QSGs) found by \citet{san65}
``resemble the quasi-stellar radio sources in many optical properties,
but they are radio quiet'' in the sense that only about one QSG per
thousand brighter than $m_{\rm pg} = 19$ appeared in the
relatively insensitive all-sky radio catalogs of the time ($S > 9$~Jy
at 178 MHz).  All of the first quasars and QSGs appear starlike
because their AGNs are far more luminous than their inferred host
galaxies, assuming their redshifts are cosmological.  The deliberately
vague name ``quasi-stellar object'' (QSO) still used today for both
quasars and QSGs was introduced by \citet{hoy66} because they doubted
the redshifts of the QSGs were cosmological and so ``at this stage we
do not think enough is known about their nature for a definitive name
to be chosen.''

Although QSGs were called ``a major new constituent of the universe''
by \citet{san65}, they are not a distinct constituent, only the
luminous tail of a continuous distribution that includes Seyfert
galaxies whose AGNs do not overwhelm their starlight.  \citet{sch83}
formally distinguished them from Seyfert galaxies by requiring QSOs contain
AGNs brighter than $M_{\rm B} = -23$, the absolute magnitude of the
brightest galaxies of stars calculated using $H_0 = 50 {\rm
  ~km~s}^{-1}{\rm ~Mpc}^{-1}$ and $q_0 = 0.1$.  In the modern flat
$\Lambda$CDM cosmology with $H_0 = 71 {\rm ~km~s}^{-1} {\rm
  ~Mpc}^{-1}$ and $\Omega_{\rm m} = 0.27$, the brightest galaxies have
$M_{\rm B} \approx -22.2$, so the spirit of the \citet{sch83}
criterion (QSOs contain AGNs brighter than any host galaxy of stars)
would change the QSO cutoff to $M_{\rm B} \approx -22.2$.  However,
like most authors, we follow the letter ($M_{\rm B} =-23$) of the
\citet{sch83} criterion.  Thus our least-luminous QSOs are actually
about twice as luminous as the brightest galaxies.  The value of this
cutoff matters because such luminous QSOs are rarely found in spiral
galaxies \citep{dun03}.

\subsection{QSOs Versus Radio Galaxies}

The cumulative radio flux-density distribution of the strongest radio
galaxies rises slightly faster than the Euclidean rate
\begin{equation}
 N(>S) = \int_S^\infty n(s) ds \propto S^{-1.5}~, 
\end{equation}
where $ N(>S)$ is the cumulative number of sources per steradian
stronger than $S$ and $n(S)dS$ is the differential number per steradian
with flux densities $S$ to $S+dS$.   If the same were
true of bright QSOs and there is one radio-loud QSO stronger than
10~Jy among 1000 optically selected QSOs, one might hope to find 30
radio sources stronger than 1 Jy and hundreds above 0.1 Jy. From this
perspective, the first directed searches for radio emission from
optically selected QSOs were surprisingly unsuccessful \citep{kel66}.
Radio sources were detected in only about 10\% of bright QSOs even
after decades of increasingly sensitive searches
\citep{kat73,str80,mil90} for sources as faint as 0.01 Jy at $\nu =
1.4{\rm ~GHz}$, or $\log[L_\nu({\rm W\,Hz}^{-1})]\approx 24$ at $z
\approx 0.2$.  These low detection rates imply that the flux-density
distribution of optically selected QSOs rises much more slowly and may
even be bimodal.  They might also indicate that QSOs and radio
galaxies do not experience the same amount of cosmological evolution
\citep{mil90}.

\subsection{Are There Two Distinct Populations of QSOs?}

\citet{str80} were the first to report a bimodal QSO flux-density
distribution, with an observed peak near $S = 1{\rm ~Jy}$ and
an inferred peak at $S < 10{\rm ~mJy}$, albeit in a very small
sample. \citet{mil90} noted a bimodal luminosity distribution implied
by the bimodal flux-density distribution of their larger sample of
QSOs in the redshift range $1.8 < z < 2.5$.  They found nine
radio sources with $\nu = 5{\rm ~GHz}$ spectral luminosities
$\log[L_\nu({\rm W\,Hz}^{-1})] > 26$, 96 upper limits below
$\log[L_\nu({\rm W\,Hz}^{-1})] \approx 25$, and nothing in between.
The term ``bimodal'' is perhaps too strong: a bimodal distribution has
two peaks.  Neither \citet{str80} or \citet{mil90} actually detected
the low-luminosity peak; its existence was only inferred from the
large number of nondetections whose radio flux densities or
luminosities would have to be spread smoothly over many orders of
magnitude to not yield a second peak.  For a recent analysis of bimodality
in light of modern optical and radio data, see \citet{bal12}.

A bimodal flux-density or luminosity distribution suggests there may
be two distinct populations of radio sources in QSOs. For example,
\citet{mil90} associated their high-luminosity peak with AGNs in
elliptical galaxies similar to the hosts of radio galaxies and the
inferred low-luminosity peak with weaker AGNs in spiral galaxies.
However, {\em any} nonuniformity or feature in these distributions is
sufficient to suggest two distinct populations.  Bimodality is
sufficient but not necessary, so we suggest that debates about
whether a clearly visible feature in a distribution makes it truly bimodal
are not relevant to the fundamental question ``Are there two distinct
populations of QSOs?''

\subsection{Further Complications}

Three factors have complicated investigations of QSO radio luminosity
functions:

(1) The radio/optical luminosity ratio of QSOs might be a better
measure of radio loudness than absolute radio luminosity alone.
\citet{sch70} proposed that the radio luminosity function ``depends on
the optical luminosity in such a manner that the ratio has a universal
distribution function'' of the form
\begin{equation}\label{req}
\Phi(F_{\rm o}, F_{\rm r}) = \Phi(F_{\rm o}) \Psi (F_{\rm r} / F_{\rm o})~.
\end{equation}
Equation~\ref{req} implies a linear correlation between radio and
optical luminosities, so the radio/optical ratio 
\begin{equation}\label{rdefeq}
R \equiv F_{\rm r} / F_{\rm o}
\end{equation}
may be a better indicator of radio activity.  \citet{pea86} countered
that almost no radio-loud QSOs are optically fainter than $M_{\rm B} =
-24$ (for $H_0 = 50 {\rm ~km~s~Mpc}^{-1}$ and $q_0 = 0.1$, equivalent
to $M_{\rm B} > -23.2$ for today's $\Lambda$CDM model), so that an
apparent correlation in the data may only reflect differences between
two populations of QSO host galaxies, optically luminous ellipticals
and optically fainter spirals.  Using FIRST \citep{bec95} direct and
statistical detections of QSOs in a very large sample, \citet{whi07}
found a clear but nonlinear correlation of radio and optical
luminosities $F_{\rm r} \propto F_{\rm o}^{0.85}$. However, Appendix C
of \citet{ive02} argues that selection effects are responsible for
such apparent correlations.

(2) Flux-density boosting in relativistic [$\gamma^2 \equiv (1 -
  v^2/c^2)^{-1} \gg 1$] radio jets may significantly broaden both the
radio flux-density and the $R$ distributions of QSOs, whose optical
emission does not appear to be strongly beamed \citep{sch79}.  Thus
radio jets in optically selected QSOs should be randomly oriented
relative to our line-of-sight, but only the small fraction $\approx
\gamma^{-2}$ of QSOs whose jets are aligned near our line-of-sight
become radio-loud quasars.  The geometry of relativistic beaming tends
to flatten the differential count of QSO radio sources 
from $n(S) \propto S^{-5/2}$ towards $n(S)
\propto S^{-4/3}$ [or $N(>S) \propto S^{-1/3}$] for intrinsically
steep-spectrum ($\alpha \equiv d \ln S / d \ln \nu \sim -1$) sources.

(3) Like the optical emission from a QSO, the radio emission is the
sum of two known components, one powered by the AGN and the other by
the host galaxy of stars.  Massive stars born within the past $\sim
10^8$ years and relativistic electrons accelerated by their supernova
remnants power the radio emission from the host galaxy, so a ``red and
dead'' elliptical galaxy with no recent star formation can be
extremely radio-quiet \citep{wal89}.  A number of features can be used
to distinguish between these two types of radio source.  The most
powerful nearby starburst is in the ultraluminous infrared galaxy Arp
220, whose spectral luminosity is $\log[L_{\rm 1.4\,GHz}({\rm
    W\,Hz}^{-1})] \approx 23.5$. 
Although the western nucleus of Arp 220 appears to contain an obscured
AGN with X-ray luminosity $L_{\rm x} \sim 10^{44}{\rm ~erg~s}^{-1}$
\citep{ran11}, the compact radio component in the western nucleus
accounts for less than 25\% of the integrated 1.4~GHz radio flux
density \citep{mun01}, so the Arp 220 starburst luminosity is not less
than $\log[L_{\rm 1.4\,GHz}({\rm W\,Hz}^{-1})] \approx 23.4$.
A starburst-powered radio source is roughly coextensive with the
starburst and is not brighter than $T_{\rm b}\sim 10^5$~K at 1.4~GHz,
has the steep spectrum $\alpha \sim -0.7$ at centimeter wavelengths
typical of optically thin synchrotron radiation, obeys the
far-infrared (FIR)/radio correlation \citep{con92}, and is not
strongly variable.  In contrast, an AGN can produce a radio source
several orders of magnitude more luminous than $\log[L_{\rm
    1.4\,GHz}({\rm W\,Hz}^{-1})]\sim 24$, its radio jets and lobes can
extend far outside the host galaxy of stars, and its radio core can be
small enough to vary significantly on time scales of years
\citep{barv05}.  The radio spectrum of an AGN can be nearly flat
($\alpha \sim 0$) over a wide frequency range, either from a
synchrotron self-absorption ``conspiracy'' among compact components
\citep{cot80} or from optically thin free-free radiation by a hot $T >
10^7$ AGN disk wind \citep{blu07}.  However, \citet{ste11} found that
``for 20 out of 22 PG quasars studied, free-free emission from a wind
emanating from the accretion disc cannot mutually explain the observed
radio and X-ray luminosity,'' eliminating free-free emission as the
cause of the radio-quiet peak in the luminosity distribution of QSOs.

Nearly all low-redshift ($z < 0.5$) radio galaxies seem to be ``red
and dead'' ellipticals whose radio sources are  powered primarily by
their AGNs.  By analogy, low-redshift QSOs whose radio sources are (1)
more luminous than $\log[L_{\rm 1.4\,GHz}({\rm W\,Hz}^{-1})]\sim 24$,
(2) have lower FIR/radio flux-density ratios than starburst galaxies,
(3) are either larger than $\sim 20 {\rm ~kpc}$ or very compact
\citep{fal96, blu98,ulv05}, (4) are strongly variable \citep{barv05}
or (5) have flat radio spectra probably reside in elliptical galaxies.
This analogy is supported by high-resolution {\it HST} images of
nearby QSO host galaxies \citep{boy98,dun03}.  QSOs close to the
original \citet{sch83} luminosity limit often obey the FIR/radio
correlation, may contain starburst-dominated radio sources
\citep[e.g.,][]{barv90,sop91,bart06,dev07}, and may be in disc or
merging galaxies \citep{boy98}. In contrast, \citet{dun03} found that
QSOs having nuclear luminosities $V < -23.5$ (assuming $H_0 = 50 {\rm
  ~km~s}^{-1} {\rm ~Mpc}^{-1}$, or $V < -22.7$ in modern $\Lambda$CDM
cosmology) are virtually all massive ellipticals with no large-scale
star formation.  Weak radio
emission correlated with the strength of the 4000~\AA~break indicates
that star formation is quite common in the host galaxies of
low-luminosity AGNs \citep{dev07}.  
We conclude that the radio properties
of QSO samples are likely to depend on the absolute magnitude cutoff,
and samples containing low-luminosity AGNs should not be used to
infer the radio properties of true QSOs.

The 1.4 GHz flux density of an ultraluminous starburst galaxy like Arp
220 would be only $S \approx 0.4{\rm ~mJy}$ at $z \approx 0.5$
and $S \approx 0.015 {\rm ~mJy}$ at $z \approx 2$, so exceptionally
sensitive observations are needed to characterize the 
starburst radio emission from individual QSO host galaxies.
\citet{kel89} detected most of the low-redshift ($z < 0.5$) Palomar
Bright Quasar Survey (PBQS) QSOs at 5~GHz and found a bimodal
spectral luminosity {\em distribution} (not {\em function}) 
peaking at $\log[L_5({\rm W\,Hz}^{-1})] \sim 26$ and $\sim 22$.  
The low-luminosity peak for QSOs spans the range $0.1 < R < 1$
with a median $\langle R \rangle \sim 0.3$.
\citet{mil93} used high-resolution radio images and
[O~III] spectra of the PBQS QSOs as evidence for a circumnuclear
starburst contribution to the low-luminosity peak.  Recent results
from significantly more sensitive Karl~G.~Jansky Very Large Array
(VLA) observations of 179 SDSS (Sloan Digital Sky Survey) QSOs in the
redshift range $0.2 < z < 0.3$ found a low-luminosity peak
\citep{kim11} 
centered on $\log[L_6({\rm W~Hz}^{-1}) \approx 22.4$ 
and $R \approx 0.3$ at $\nu = 6$~GHz.
  The radio sources contributing to this peak have a median spectral
  index $\alpha = -0.7$ consistent with starburst radio emission.
  Unfortunately their median flux densities are only $\langle S
  \rangle \approx 0.13$~mJy at 6~GHz and $\langle S \rangle \approx
  0.33$~mJy at 1.4~GHz, so making follow-up observations will be
  difficult. The \citet{kim11} radio luminosity function suggests that
  most radio sources with $\log[L_6({\rm W~Hz}^{-1})] > 23$ at 6~GHz
(or $\log[L_{1.4}({\rm W~Hz}^{-1}) > 23.4$ at 1.4~GHz) in
    low-redshift QSOs are powered primarily by AGNs, not starbursts.

\citet{kel94} reobserved ``radio quiet'' QSOs and low-luminosity AGNs
from their BQS sample \citep{kel89} at 5~GHz with $18''$ and
$0\,\farcs5$ resolution and resolved a number of compact cores and
extended lobes similar to those in radio-loud quasars, suggesting that
central engines exist in galaxies and quasars over a wide range of
radio luminosity.  They noted, however, that their resolved ``radio
quiet'' QSOs had radio luminosities well in excess of that expected
from the underlying galaxy.

Several high-resolution radio observations have demonstrated that most
mJy-level radio sources in QSOs contain radio components so compact
that they must be AGN-powered.  Three flat-spectrum ($\alpha > -0.5$)
``radio intermediate'' (defined by $10 < R < 250$ at $\nu = 5$ GHz)
QSOs were found to have brightness temperatures $T_{\rm b} > 10^9$~K
\citep{fal96}, much too high for starburst radio emission.  However,
only one of them (PG 1309+355 at $z = 0.184$) satisfies our $M_{\rm B}
< -23$ criterion for a QSO, and its 5 GHz radio luminosity
$\log[L_5({\rm W~Hz}^{-1})] \approx 24.6$ is an order of magnitude
higher than the most luminous starburst yet found in a low-redshift
QSO.  \citet{blu98} imaged 12 ``radio-quiet quasars'' with the VLBA
and found strong evidence for jet-producing central engines in eight
of them.  However, four of the ``quasars'' are low-luminosity AGNs
with $M_{\rm B} > -23$ (Mrk 335, III Zw 2, Mrk 705, and II Zw 171).
Five of the eight true QSOs were detected by the VLBA, but all five
have radio luminosities in the range $23.8 < \log[L_{1.4}({\rm
    W~Hz}^{-1}) < 26.2]$ high enough that we expect a significant AGN
contribution.  \citet{ulv05} considered a sample of 11 ``radio-quiet
quasars'' having total 5 GHz flux densities of 4 mJy or greater.
Three have no compact cores stronger than 1 mJy and were discarded
from the sample.  New VLBA images were made of five, one of which (UGC
09412) is a low-luminosity AGN.  Two of the true quasars are in the
\citet{blu98} sample above.  The remaining QSOs (J0804+6459 at $z =
0.148$ and UM 275 at $z = 2.137$) have radio luminosities
$\log[L_5({\rm W~Hz}^{-1})] = 23.8$ and $26.0$, respectively, and 5
GHz radio/optical flux ratios $R = 3$ and 22, respectively, much
higher than we expect for radio sources powered only by starbursts.
We believe their conclusion ``Thus, the most likely explanation for
their radio properties is simply the radio-quiet quasars are similar
to their 'traditional' radio-loud cousins but have less powerful radio
jets.'' \citep{ulv05} applies only to QSOs containing radio sources
stronger than starbursts produce.  We conclude that, unfortunately,
VLB observations are simply not sensitive enough ($\sigma \ll 0.1 {\rm
  ~mJy}$) to address the question of starburst radio emission in large
samples of radio-quiet quasars.

Finally, \citet{barv05} compared radio variability in samples of
radio-quiet (defined by having $R < 3$), radio-intermediate, and
radio-loud quasars.  They found that radio variability is independent
of radio loudness.  Of the 11 ``radio-quiet quasars'' monitored, four
are low-luminosity AGNs (Mrk 335, Mrk 1148, UGC 11763, and PGC
070504). Four of the seven true QSOs are variable radio sources and
three are not.  Among the variable QSOs, only PG 0052+251 at $z =
0.155$ is sufficiently radio quiet ($\log[L_8({\rm W~Hz}^{-1})] =
22.6$ and $R \sim 0.3$) that a starburst might have contributed
significantly to its radio emission.  This single source is consistent
with the 6 GHz radio luminosity functions shown in Figure 6 of
\citet{kim11}, which suggests that about one in four QSO radio sources
with $\log[L_6({\rm W~Hz}^{-1})] = 22.6$ should be AGN-dominated.
We believe their conclusion ``...the radio emission from
radio-quiet quasars originates in a compact structure intimately
associated with the active nucleus.  The alternative hypothesis, that
the emission from radio-weak quasars is from a starburst, is ruled
out.'' \citep{barv05} has been justified only for radio sources stronger
than $\log[L_8({\rm W~Hz})] \sim 24$ and $R > 1$ in QSOs, not for
the weaker radio sources expected from starbursts.

\subsection{Outline of This Paper}

QSOs have such broad radio luminosity functions that their radio
emission can be characterized only via large
unbiased samples of radio-detected QSOs. Sensitive directed searches
for radio emission from QSOs have been impeded by the small numbers of
radio detections.  A typical VLA observing program might target $N \sim
100$ QSOs.  Detecting statistically useful numbers $N \sim 100 \gg
N^{1/2}$ of radio sources requires either a detection rate near 100\%
or a much larger QSO sample if the detection rate is low (e.g., $N
\sim 1000$ if the detection rate is only $\sim 10$\%).

Furthermore, nearby QSOs in which starburst-powered radio sources can
be detected individually have such a low sky surface density that a
statistically useful sample can be obtained only
from an optical survey covering more than a steradian
with uniformly accurate multicolor photometry.  The SDSS DR7 (Data
Release 7) QSO sample \citep{sch10} covers $\Omega \approx 2.66$ sr at
Galactic latitudes $\vert b \vert > 30^\circ$ and satisfies that
requirement.  This paper is based on NRAO VLA Sky Survey (NVSS)
\citep{con98} 1.4 GHz flux densities of (1) a volume-limited sample of
1313 DR7 QSOs with $0.2 < z < 0.45$ (Sec.~\ref{lowzsec}) and (2) a
magnitude-limited sample of 2471 DR7 QSOs in the redshift range $1.8 <
z < 2.5$ (Sec.~\ref{hizsec}).

Above the NVSS catalog limit $S \approx 2.4$~mJy~beam$^{-1}$, the
detection rate in each sample is only about 10\%, but the QSO samples
are large enough to yield hundreds of NVSS detections.  In addition,
most of the remaining QSOs can be detected statistically on the NVSS
images at levels $S\sim 0.1$~mJy~beam$^{-1}$.  The main result of the
direct detections is that the differential luminosity function
$\rho_{\rm m}(L)$ of strong radio sources powered by the AGNs in QSOs
is an extremely flat power law, so flat that extrapolating to lower
luminosities predicts essentially no statistical detections at the $S
\sim 0.1$~mJy level.  Our high statistical detection rate implies a
``bump'' in the radio luminosity function of QSOs in the range
$\log[L({\rm W\,Hz}^{-1})] \sim 23$ to 24, depending on redshift.  We
attribute this bump to radio emission powered by luminous starbursts
in the host galaxies of {\it most} bright QSOs
(Sec.~\ref{qsolumfsec}).

\section{The Volume-limited $0.2 < z < 0.45$ QSO Sample}\label{lowzsec}

The SDSS DR7 QSO catalog \citep{sch10} is complete to $i = 19.1 {\rm
  ~mag}$ over a solid angle $\Omega = 2.66 {\rm~sr}$ around the North
Galactic Pole.  It contains the small sample of 179 color-selected
QSOs defined by $M_{\rm i} < -23$ in the narrow redshift range $0.2 <
z < 0.3$ studied by \citet{kim11} and the larger sample of 1313 QSOs
in the wider redshift range $0.2 < z < 0.45$ discussed here.  Note that
these magnitudes were calculated for an $H_0 = 71 {\rm
  ~km~s~Mpc}^{-1}$ $\Lambda$CDM cosmology, so our QSO cutoff
corresponds to $M_{\rm B} \approx -23.8$ for comparison with
\citet{sch83} and should discriminate against the low-luminosity QSOs
that can appear in spiral galaxies \citep{dun03}.  Both samples are
strictly volume limited because any $M_{\rm i} < -23$ QSO with $z <
0.45$ is brighter than $i = 19 {\rm ~mag}$.  There are no more
color-selected QSOs with $0.2 < z < 0.45$ to be found in that 21\% of
the sky, so no observation 
made at this time and place in the universe
can reduce our
statistical errors by more than about a factor of two---our sample
variance is approaching the ``cosmic variance'' limit set by the
finite observable volume of the universe \citep{kam97}.

\subsection{Sources in the NVSS Catalog}

The entire DR7 area is covered by the NVSS, whose source catalog is
complete for statistical purposes above a peak flux density $S_{\rm p}
\approx 2.4 {\rm ~mJy~beam}^{-1}$ at 1.4 GHz.  In the redshift range
$0.2 < z < 0.45$ the 45~arcsec FWHM (full width between half-maximum
points) beam of the NVSS spans 150 to 250 kpc, more than enough to
cover the host galaxy of stars and most of the extended radio emission
powered by an AGN.  
Consequently the NVSS sees most QSOs as point sources and the NVSS
integrated flux-density limit is $S \approx S_{\rm p} = 2.4 {\rm
  ~mJy}$.
For radio sources with spectral index $\alpha \approx -0.7$, the
corresponding $\nu = 1.4$~GHz (in the source frame) NVSS
spectral-luminosity detection limits range from $\log[L_\nu({\rm
    W\,Hz}^{-1})] = 23.4$ at $z = 0.2$ to $24.2$ at $z = 0.45$.  Thus
most of these sources are too luminous to be powered primarily by
starbursts.

\begin{figure}
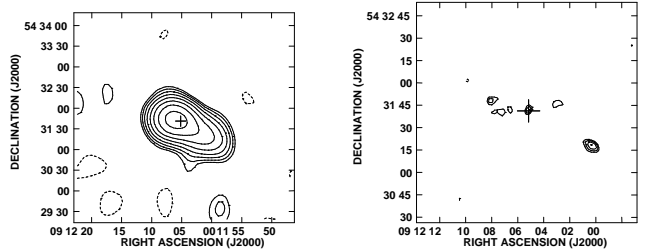

\includegraphics[width=1.6in]{f1left.eps}\hfill
\includegraphics[width=1.6in]{f1right.eps}
\caption{NVSS (left panel) and FIRST (right panel) images of SDSS
  J091205.16+543141.2 (cross).  Contours are separated by factors of
  $\pm 2^{1/2}$ starting at 1 mJy beam$^{-1}$ (NVSS) and 0.5 mJy
  beam$^{-1}$ (FIRST).  \label{0912idfig}}
\end{figure}

To ensure completeness even for radio sources larger than the NVSS
beam, we examined all NVSS catalog entries within 120 arcsec of each
QSO.  We rejected components centered more than a few arcsec from the
optical position unless their morphologies in the NVSS and/or
higher-resolution (5.4~arcsec FWHM) FIRST \citep{bec95} images
plausibly link them with the QSO.  Unresolved or symmetric NVSS
sources are easily identified with QSOs by radio-optical position
coincidence alone, but asymmetric multicomponent sources can be
difficult.  For example, the left panel of Figure~\ref{0912idfig}
shows two marginally resolved NVSS components whose centers straddle
the optical position (cross) of SDSS J091205.16+543141.2.  Two lobes
and a core are clearly resolved in the FIRST image (right panel),
confirming the otherwise uncertain NVSS identification.  Because the
NVSS has low resolution, it has high surface-brightness sensitivity.
The two NVSS components contain essentially all of the QSO flux
density, 23.8 mJy, while the three FIRST components add up to only 6.6
mJy.  This example illustrates how the NVSS and FIRST complement each
other for identifying extended radio sources and measuring their flux
densities.  

We also found that the FIRST peak flux densities of QSOs fainter than
$S \approx 10 {\rm ~mJy}$ are usually close to the NVSS flux
densities, and very few faint NVSS sources are not in the FIRST
catalog.  This indicates that most of the fainter radio sources
produced by these QSOs have half-power linear sizes $ \lesssim 5 {\rm
  ~kpc}$ and the likelihood that the NVSS has resolved out any faint
sources is small.  Unfortunately, radio sources powered by starbursts
and and by low-luminosity jets from AGNs are both consistent with this
size limit; sub-arcsecond resolution is needed to distinguish between
them \citep{con91}.  FIRST probably does not resolve out many sources
in the $1 < S{\rm (mJy)} < 2.4$ range either, but we have not included
the FIRST flux densities for these sources in our statistical analysis
because they are systematically low by about 0.25~mJy owing to
``snapshot'' bias \citep{whi07}.  Also the FIRST source count
\citep{whi97} indicates a rapid fall in differential
completeness below $S \approx 2$~mJy.  Although the NVSS image noise
is higher, the snapshot bias is lower, so we chose to study all
sources fainter 2.4~mJy statistically via their amplitude distribution
on NVSS images 
(see Section~\ref{pofdsubsec} for a complete discussion of snapshot
bias in FIRST and NVSS) instead.

A number of stronger sources are clearly resolved by the NVSS, but
most have brightnesses well above the $2.4{\rm ~mJy~beam}^{-1}$ NVSS
catalog limit.  Thus the NVSS integrated flux densities of radio-loud
QSOs should be accurate and the NVSS should not have missed any
radio-loud QSOs.  There are 163 NVSS detections with $0.2 < z < 0.45$
(Table~\ref{nvsstable}), of which only 37 are in the narrow redshift
range $0.2 < z < 0.3$ that was also covered by our more sensitive
directed VLA observations at 6 GHz \citep{kim11}.

\subsection{The 1.4~GHz Flux-density Distribution of  
QSOs Stronger Than 2.4~mJy}\label{countsec}

\begin{figure}
\includegraphics[width=\linewidth,trim=20 70 60 100,clip ]{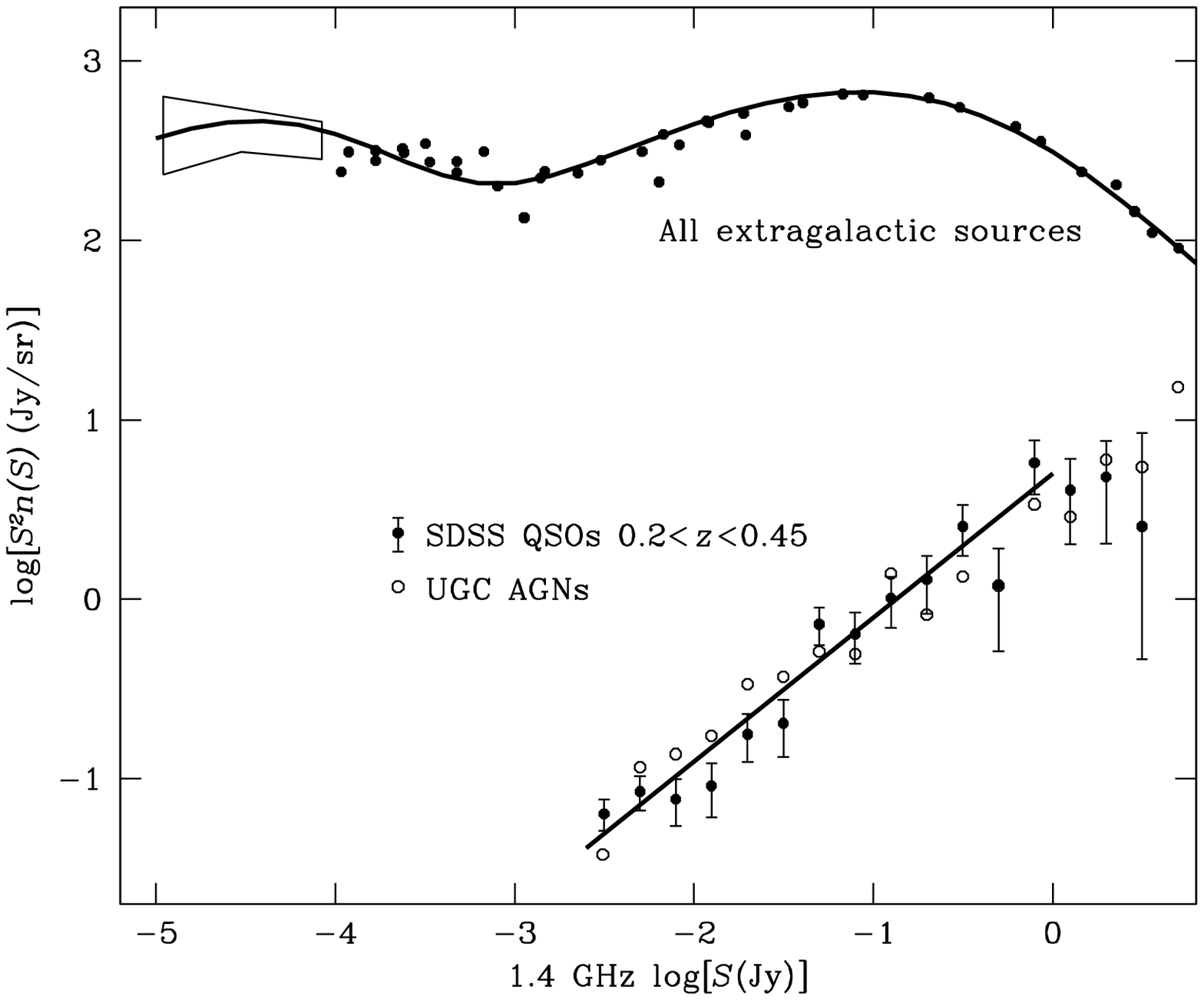}
\caption{The peaks in the brightness-weighted count $S^2 n(S)$ of all
  extragalactic sources \citep{con84,mit85} near $\log[S({\rm Jy})] =
  -1$ and $\log[S({\rm Jy})] = -4$ match the expected contributions
  from AGNs (primarily in radio galaxies) and star-forming galaxies,
  respectively, at typical redshifts $z \sim 1$.  The low-redshift
  SDSS QSO count (filled points with error bars) is well fit by the
  power-law (solid line) whose slope also matches the count (open
  circles) of nearby ($z \lesssim 0.05$) UGC galaxies whose radio
  sources are powered primarily by AGNs \citep{con02}.  All points in
  this figure are on the same ordinate scale, so the vertical
  separation between the points for all extragalactic sources near the
  top and SDSS QSOs with $0.2 < z < 0.45$ near the bottom shows how
  the fraction of all radio sources that are in low-redshift QSOs
  drops from about 1\% near $S = 1$~Jy to about 0.01\% near $S = 1$
  mJy.  The nearly equal densities of SDSS QSOs and UGC galaxies is
  only a coincidence.  Abscissa: log 1.4 GHz flux density (Jy).
  Ordinate: Differential source count $n(S)$ multiplied by $S^2$
  (Jy~sr$^{-1}$). \label{s2nfig}}
\end{figure}

Let $n(S)dS$ be the differential number of sources per steradian in
the flux-density range $S$ to $S+dS$. Then the brightness-weighted
count $S^2n(S)$ is proportional to the contribution $d T_{\rm b}$ of
sources in each logarithmic flux-density range to the sky background
temperature:
\begin{equation}
\biggl[ {d T_{\rm b} \over d \log(S)}\biggr] = 
\biggl[ { \ln(10) c^2 \over 2 k_{\rm B} \nu^2} \biggr] S^2 n(S)~.
\end{equation} 
The brightness-weighted count of all extragalactic sources at 1.4 GHz
(mostly radio galaxies) is shown by the upper curve in
Figure~\ref{s2nfig}.  The lower curve in Figure~\ref{s2nfig} shows that
(1) the 1.4 GHz flux-density distribution of the $0.2 < z < 0.45$ QSOs
is a smooth power law above 2.4~mJy and (2) the QSO detection rate
grows very slowly with improvements in sensitivity.  In the
flux-density range $2.4 < S{\rm (mJy)} < 1000$, the unbinned
maximum-likelihood power-law fit \citep{cra70} to the
brightness-weighted differential source count of these QSOs is
$S^2n(S) = (5.0 \pm 0.4) \times S^{\,0.80 \pm 0.02} {\rm
  ~Jy~sr}^{-1}$; that is,
\begin{equation}
n(S) \propto S^{-\beta}~,
\end{equation}
where $\beta = 1.20 \pm 0.02$ is much lower than the static Euclidean
$\beta = 2.5$.  The Kolmogorov-Smirnov (K-S) test for goodness of fit
yields the probability $P > 0.33$ that the source count is 
consistent with the power-law fit.  There are no gaps or other features
in this QSO flux-density distribution to suggest more than one source
population above $S = 2.4$~mJy, where we expect primarily AGN-dominated
radio emission.

Because the flux-density range spanned by these QSOs is large, it
is also useful to consider the number $n_{10}(S)$ of radio sources in each
{\it logarithmic} flux-density interval centered on $S$.  We define $n_{10}$ by
\begin{equation}
n_{10}(S) d \log S = n(S) dS
\end{equation}
so that
\begin{equation}\label{nmeq}
n_{10} (S)= \ln(10) S\, n(S) \propto S^{1-\beta}.
\end{equation}
For our $0.2 < z < 0.45$ QSOs, $(1-\beta) = -0.20 \pm 0.02$. The QSO
counts are ``flat'' in the sense that $n_{10}(S)$ is nearly
independent of $S$.  In each successively lower decade of flux
density, the expected number of radio detections rises by only ${\rm
  dex}(0.2) - 1 \approx 60$\%.  This is quite different from the
``steep'' counts of radio galaxies or of all sources in a static
Euclidean universe, where $(1-\beta) \sim -1.5$ regardless of the
luminosity function, and the number of radio detections should
increase by about $3000$\% per decade.

The brightness-weighted count of our $0.2 < z < 0.45$ QSOs clearly
peaks at the highest flux densities (Figure~\ref{s2nfig}).  The
strongest 12 radio sources (those with $S > 900$\,mJy) comprise only
1\% of the 1313 QSOs but contribute more than half of the total radio
flux density from the whole sample, even if all 1148 radio sources not
in the NVSS catalog had flux densities equal to their $5 \sigma =
2.4$\,mJy NVSS upper limits.

\subsection{The 1.4 GHz Luminosity Function and its Evolution, 
$0.2 < z < 0.45$}\label{lumfsubsec}

\begin{figure}
\includegraphics[width=\linewidth,trim=30 40 100 120,clip ]{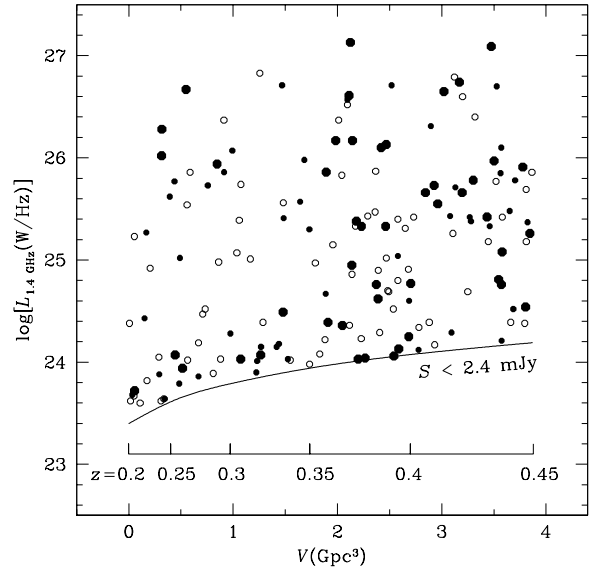}
\caption{The 1.4 GHz luminosity distribution of QSOs detected by the
NVSS is shown as a function of the volume enclosed between the
redshift of each QSO and the sample's lower redshift limit $z = 0.2$.
The $S = 2.4 {\rm ~mJy}$ NVSS sensitivity limit is indicated by the
curve. The QSO sample is volume limited, so the density of points
above the curve is proportional to its 1.4 GHz luminosity function.
Symbols distinguish QSOs by their absolute magnitudes: large filled
symbols for $M_{\rm i} < -24$, small filled symbols for $-24 < M_{\rm
  i} < -23.5$, and open symbols for $-23.5 < M_{\rm i} <
23$. Abscissa: comoving volume $V$ (Gpc$^3$).  Ordinate: log 1.4 GHz
spectral luminosity.  Parameter: redshift $z$. \label{lovfig}}
\end{figure}

The 1.4 GHz spectral luminosity distribution of QSOs in our sample is
shown in Figure~\ref{lovfig} as a function of the comoving volume in
$\Omega = 2.66 {\rm ~sr}$ between the redshift cutoff at $z = 0.2$ and
the redshift of each QSO.  The QSO sample is strictly volume limited
and the radio detections are complete above the $S = 2.4 {\rm ~mJy}$
NVSS limit indicated by the curved line.  Consequently the plotted density
of points above the curved line is proportional to the 1.4~GHz
spectral luminosity function
\begin{equation}
\rho_{\rm m} (L_\nu) \equiv \ln(m) L_\nu \rho(L_\nu)~,
\end{equation}
where $m \equiv {\rm dex}(0.4) = 1$ ``magnitude'' and
$\rho(L_\nu)dL_\nu$ is the comoving space density of sources in the
spectral luminosity range $L_\nu$ and $L_\nu + dL_\nu$.  The QSOs in
Figure~\ref{lovfig} have been grouped into three absolute-magnitude bins: $M_{\rm i} < -24$ (large filled symbols), $-24 < M_{\rm i}
< -23.5$ (small filled symbols), and $-23.5 < M_{\rm i} < -23$ (open
symbols).

Figure~\ref{lovfig} shows that: (1) The 1.4 GHz spectral luminosity
function has no statistically significant features (regions of
significantly low or high density) in the sampled range $24 <
\log[L_{\rm 1.4~GHz}({\rm W\,Hz}^{-1})] < 27$ at any redshift $0.2 < z
< 0.45$. (2) The radio and optical luminosities do not appear to be
correlated.  However, the radio luminosities span three decades, while
only five of the 163 QSOs are brighter than $M_{\rm i} = -25.5$, one
decade brighter than the $M_{\rm i} = -23$ optical cutoff, so a linear
correlation would be difficult to discern even if there were one.  The
fact that a linear correlation would be difficult to discern suggests
that studying radio/optical ratios $R$ \citep{sch70} instead of flux
densities $S$ is not likely to yield new insights into low-redshift
optically selected QSOs.  We note that \citet{mah12} also found no
correlation of radio and optical luminosities in their sample of
low-luminosity X-ray selected QSOs.  (3) The plotted density of points
increases to the right, indicating radio density evolution.

\begin{figure}
\includegraphics[width=\linewidth,trim=20 40 100 180,clip ]{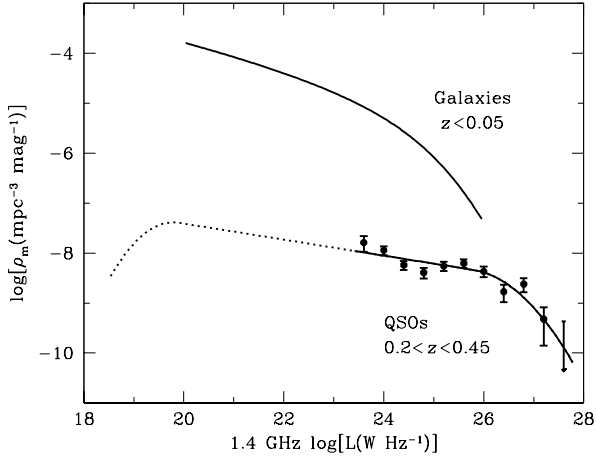}
\caption{The 1.4 GHz luminosity function of QSOs in the redshift range
$0.2 < z < 0.45$ based on NVSS detections is shown by data points at
$\log[L_{\rm 1.4\,GHz}({\rm W\,Hz}^{-1})] = 23.6$ to 27.2 by 0.4 and
an upper limit at 27.6.  The continuous curve approximating the QSO
data follows equations~\ref{allzlumfeq} and~\ref{allzlumftrunceq}.
The dotted line is a power-low extrapolation to lower luminosities,
cut off below $\log[L_{\rm 1.4~GHz}({\rm W\,Hz}^{-1})] \sim 20$ to
ensure that the number of radio sources equals the number of QSOs.
Abscissa: log 1.4 GHz spectral luminosity (W Hz)$^{-1}$.  Ordinate:
log comoving space density (Mpc$^{-3}$) of sources per ``magnitude'' =
dex(0.4) luminosity range. \label{lumffig}}
\end{figure}

We calculated the 1.4 GHz spectral luminosity function of
color-selected QSOs brighter than $M_{\rm i} = -23$ in the redshift
range $0.2 < z < 0.45$ using the standard $1 / V_{\rm max}$ method.
If there are $N$ sources in the logarithmic bin of width $m$
centered on spectral luminosity $L_\nu$, then
\begin{equation}
\rho_{\rm m}(L_\nu) = \sum_{i=1}^N \biggl( {1 \over V_{\rm max}}\biggr)_i~,
\end{equation}
where $V_{\rm max}$ is the comoving volume in the sample solid angle
$\Omega = 2.66{\rm ~sr}$ between the minimum redshift $z = 0.2$ and the
maximum redshift $z_{\rm max} \leq 0.45$ at which the source could
have been detected.  The rms statistical uncertainty of $\rho_{\rm
  m}(L_\nu)$ is
\begin{equation}
\sigma = \biggl[\sum_{i=1}^N \biggl({1 \over V_{\rm max}}\biggr)^2_i
\biggr]^{1/2}~.
\end{equation}
The binned luminosity function of QSOs averaged over the redshift
range $0.2 < z < 0.45$ is shown in Figure~\ref{lumffig}.  The data
points can be fit by the power law
\begin{eqnarray}\label{allzlumfeq}
\log[\rho_{\rm m}({\rm mag}^{-1}{\rm~Mpc}^{-3})] = -4.21
\nonumber \\
-0.16 \log[L_{\rm 1.4\,GHz}({\rm W\,Hz}^{-1})] 
\end{eqnarray}
in the luminosity range $23.4 < \log[L_{\rm 1.4\,GHz}({\rm
    W\,Hz}^{-1})] < 25.8$.  The power-law slope
\begin{equation}
\epsilon \equiv d\log(\rho_{\rm m}) / d \log(L_\nu) = -0.16
\end{equation}
is close to the slope $(1-\beta) \approx -0.2$ of $n_{10}(S)$, as
suggested in Section~\ref{countsec}.  It is somewhat flatter than the
luminosity function of UGC radio galaxies powered by AGNs (upper curve
in Figure~\ref{lumffig}).

At higher luminosities, the luminosity function must fall below this
flat power law; otherwise the spectral power density contributed by
the most luminous sources would diverge.  The amount
$\Delta\log(\rho_{\rm m})$ by which the observed luminosity function
falls below the power-law fit can be approximated by the (somewhat
arbitrary, but simple and yielding a satisfactory fit to the limited
data) quadratic form indicated by the lower continuous curve in
Figure~\ref{lumffig}:
\begin{eqnarray}\label{allzlumftrunceq}
\Delta\log[\rho_{\rm m}({\rm mag}^{-1}{\rm\,Mpc}^{-3})] = 
\nonumber \\
-\bigl[\bigl(\log[L_{\rm 1.4\,GHz}({\rm W\,Hz}^{-1})] - 25.8\bigr)/1.6\bigr]^2~
\end{eqnarray}
when $\log[L_{\rm 1.4\,GHz}({\rm W\,Hz}^{-1})] > 25.8$.  Finally, the
total number of optically selected QSOs is finite ($N = 1313$ over
$\Omega = 2.66{\rm ~sr}$ in the redshift range $0.2 < z < 0.45$), so
the total number of radio sources predicted by the luminosity function
should not exceed this number.  
Any extrapolation of the fitted luminosity function to lower
luminosities (dotted line in Figure~\ref{lumffig}) must cut off below
$\log[L_{\rm 1.4~GHz}({\rm W~Hz}^{-1})] \sim 20$ prevent the total
number of radio sources from exceeding the number of QSOs.

The NVSS catalog should contain all sources with spectral luminosities
$\log[L_{\rm 1.4\,GHz}({\rm W~Hz}^{-1})] > 24.2$ throughout the
redshift range $0.2 < z < 0.45$.  Low-redshift radio sources this
strong are probably dominated by emission from AGNs, not starbursts.
The evolution of their luminosity function with redshift may be
characterized in terms of pure density evolution (shifting the
luminosity function horizontally in Figure~\ref{lumffig})
\begin{equation}
\log[\rho_{\rm m}(L, z)] =  \log[g(z) \rho_{\rm m}(L, 0)]
\end{equation}
or pure luminosity evolution (shifting the luminosity function
vertically in Figure~\ref{lumffig})
\begin{equation}
\log[\rho_{\rm m}(L, z)] = \log[\rho_{\rm m}(L/f(z), 0)]~,
\end{equation}
where the functions $g(z)$ and $f(z)$ specify the amounts of density
and luminosity evolution, respectively.  
For power-law density evolution $g(z) = (1+z)^\delta$, $\delta = +6.0
\pm 1.8$ is needed to match the 1.4~GHz data in this portion of the
redshift--luminosity plane, and the fit is satisfactory [$P(\chi^2) >
  0.2$]. Thus the evolving 1.4 GHz luminosity function of QSOs in the
redshift range $0.2 < z < 0.45$ is well approximated by replacing
Equation~\ref{allzlumfeq} with
\begin{eqnarray}\label{allzevlumfeq}
\log[\rho_{\rm m}({\rm mag}^{-1}{\rm\,Mpc}^{-3})] = - 5.06
\nonumber \\
-0.16 \log[L_{\rm 1.4\,GHz}({\rm W\,Hz}^{-1})] + 6.0\log(1+z)
\end{eqnarray}
and leaving the cutoff Equation~\ref{allzlumftrunceq} unchanged.  

Pure luminosity evolution of the form $f(z) =
(1+z)^\lambda$ is indistinguishable from pure density evolution in the
power-law portion of a luminosity function.  However, the small
power-law slope $\epsilon \approx -0.16$ of the QSO luminosity
function means that very strong luminosity evolution is needed to have
the same effect as moderate density evolution: $\lambda \sim -\delta /
\epsilon \sim 38$. The high-luminosity cutoff cannot evolve this
rapidly because there are only two sources with $\log[L_{\rm
    1.4\,GHz}({\rm W\,Hz}^{-1})] > 26$, so pure luminosity evolution
does not fit our radio data.

\subsection{Why is the QSO Flux-density Distribution So Flat?}

The flat flux-density distribution $n_{10}(S)$ of radio sources in low-redshift
optically selected QSOs does {\it not} prove that they are
significantly different from the radio sources in nearby optically
selected galaxies.  The open circles in Figure~\ref{s2nfig} show the
weighted count $S^2n(S)$ of AGN-dominated radio sources in
low-redshift ($z \lesssim 0.05$) UGC galaxies \citep{con02}; their
power-law slope is almost the same. (Their sky density is also nearly
the same, but that is just a coincidence that depends on the size of
the UGC sample.)

Why do these optically selected QSO and galaxy samples have such flat
1.4 GHz flux-density distributions? Consider the flux-density
distributions in two extreme cases: (1) In an infinite static
Euclidean universe, the slope of $n_{10}(S)$ would be $1-\beta =
-1.5$, for any luminosity function.  (2) If all of the sources in a
sample were at exactly the same distance, the slope of $n_{10}(S)$
would equal the slope of the luminosity function $\rho_{\rm m}(L)$.
For the more realistic case of a sample of sources spanning a finite
redshift range $z_{\rm min}$ to $z_{\rm max}$, the flux-density
distribution will be the convolution of the luminosity function with a
smoothing function whose width is roughly $(z_{\rm max} / z_{\rm
  min})^2$, the range of flux densities for a source of fixed
luminosity as it is moved across the redshift range.

For both the SDSS QSO and UGC galaxy samples, the power-law portions
of their 1.4 GHz luminosity functions (Figure~\ref{lumffig}) are much
wider than $(z_{\rm max} / z_{\rm min})^2 \sim 6$ for steep-spectrum
QSOs confined to $0.2 < z < 0.45$.  About $7/8$ of the volume in $0 <
z < 0.05$ containing the AGN-dominated UGC galaxies is in the narrow
redshift range $0.025 < z < 0.05$ for which the smoothing width is
only $\approx 4$.  In the small-smoothing limit, the power-law slope
$(1-\beta)$ of $n_{10}$ naturally approaches the power-law slope $d
\log(\rho_{10}) / d \log L_\nu$ of the differential luminosity
function expressed as a source density per log radio luminosity.  Thus
the flat and nearly equal flux-density distributions of both the SDSS
QSO and the AGN-dominated UGC galaxy samples are telling us only that
both samples have relatively broad and flat radio luminosity
functions.  If the AGN-powered radio sources in both optically
selected samples are intrinsically similar, we should {\it expect} our
QSOs to have a distinctly sub-Euclidean flux-density distribution.

The observed $\beta \approx 1.2$ is also consistent with relativistic
beaming, which tries to make $\beta \approx 4/3$ \citep{sch79}.
However, relativistic beaming of the modestly luminous \{typical
$\log[L_{\rm 1.4\,GHz}(\rm W\,Hz)^{-1}] \sim 25$\} NVSS radio sources
in our low-redshift QSO sample is probably insufficient to influence
the radio flux-density distribuion.  In beaming models, the most
luminous sources are the most highly beamed, so that most of their
flux densities arise in highly relativistic ($\gamma \sim 5$--10)
components approaching within $\sim \gamma^{-1}$ radians of our line of
sight.  Relativistic jets are compact and one-sided, but the most
luminous sources in our sample are not. Table~\ref{nvsstable} lists 26
sources with (apparent) $\log[L_{\rm 1.4\,GHz}({\rm W\,Hz}^{-1})] >
26$, but only seven have even half of their flux density unresolved by
FIRST; the rest are too extended (linear diameter $\gtrsim 5{\rm
  ~kpc}$) and centered on their optical QSO positions to be strongly
boosted.  The seven fairly compact sources contribute only 34\% of the
total apparent luminosity of the 26 luminous sources, so removing them
would reduce the total luminosity of the sample by at most $\Delta
\log(L_\nu) = -0.18$.  What we see is what we get, and we can
calculate reasonably accurate intrinsic luminosities and luminosity
functions by assuming isotropic radio emission.

\subsection{NVSS Statistical Detections: Numerous Faint Sources 
Powered by Star Formation?}\label{pofdsubsec}

To detect radio sources weaker than $S = 2.4$\,mJy, we originally
measured NVSS peak flux densities at the positions of all $N = 179$
color-selected QSOs in the narrow redshift band $0.2 < z < 0.3$
\citep{kim11}.  The NVSS image noise has a nearly Gaussian amplitude
distribution with rms $\sigma = 0.45 \pm 0.01 {\rm ~mJy~beam}^{-1}$
and zero mean \citep{con98}.  A starburst with luminosity
$\log(L/L_\sun) \approx 11$ obeying the FIR/radio correlation should
produce a radio source with spectral index $\alpha \approx -0.7$ and
spectral luminosity $\log[L_\nu({\rm W\,Hz}^{-1})] \approx 22.5$ at
$\nu = 1.4$~GHz in the source frame.  Such a source would yield an
observed 1.4 GHz flux density ranging from $S = 0.12{\rm ~mJy}$ at $z
= 0.3$ to $S = 0.30{\rm ~mJy}$ at $z = 0.2$.  The NVSS is not
sensitive enought to detect such faint sources individually, but the
distribution of peak flux densities on QSO positions can be used to
detect a large ($N > 100$) sample of them statistically.

\citet{whi07} discovered a potential problem with statistical
detections: the nonlinear CLEAN bias associated with the high
dirty-beam sidelobes of snapshot surveys like FIRST and NVSS can
produce a ``snapshot bias'' that lowers the image peak flux densities
of sources fainter than the catalog limit.  The snapshot
bias for a faint source is about 40\% of the peak flux density on a FIRST
image.  To see if snapshot bias might affect the peak flux densities
of faint QSOs on NVSS images, we followed their technique and measured
NVSS peak flux densities on the positions of faint sources found in
the sensitive ($\sigma \approx 0.023 {\rm ~mJy~beam}^{-1}$) 1.4 GHz
VLA image covering the Spitzer First-Look Survey \citep{con03}.  This
image was made from data having very good $(u,v)$-plane coverage, so
its snapshot bias should be extremely low. The results of this
comparison are listed in Table~\ref{biastable}.  They show that
the NVSS snapshot bias is too small to measure over the flux-density
range $0.115 < S_{\rm p} < 0.460 {\rm ~mJy\,beam}^{-1}$.

\begin{figure}
\includegraphics[width=\linewidth,trim=50 30 40 100,clip ]{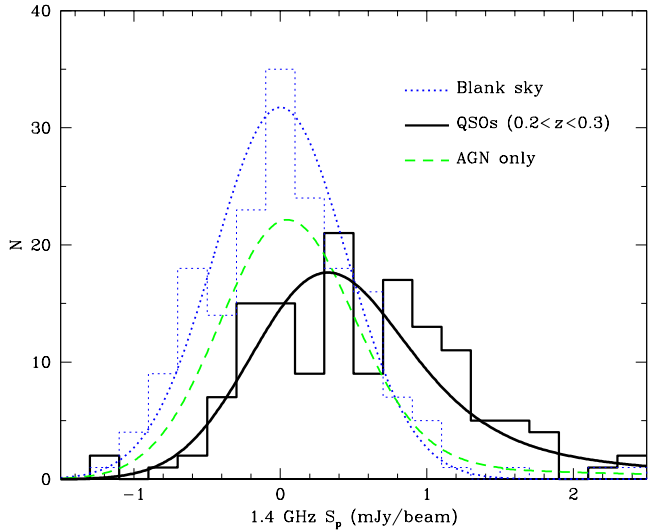}
\caption{The continuous histogram shows the distribution of NVSS peak
  flux densities for the volume-limited sample 179 SDSS QSOs in the
  redshift range $0.2 < z < 0.3$.  The dotted histogram (colored blue
  in the online version) indicates the distribution of NVSS peak flux
  densities at ``blank sky'' positions one degree north of the QSOs,
  and it is well fit by a Gaussian of mean $\langle S_{\rm p}\rangle =
  -0.01 \pm 0.04 {\rm ~mJy~beam}^{-1}$ and rms $\sigma =0.48 \pm 0.04
  {\rm ~mJy~beam}^{-1}$ (dotted curve, colored blue in the online
  version). The dashed curve (colored green in the online version) is
  the distribution predicted by extrapolating the observed luminosity
  function of SDSS QSOs to lower luminosities, and the continuous black
  curve represents the distribution predicted by our best-fit model in
  which both AGNs and star-forming host galaxies contribute to the
  radio emission of QSOs. \label{pofdfig}}
\end{figure}

Figure~\ref{pofdfig} shows the distribution of NVSS peak flux
densities $S_{\rm p}$ at the optical positions of the 179 QSOs with
$0.2 < z < 0.3$ (continuous histogram) and at the 179 ``blank sky''
positions exactly one degree north of the QSOs (dotted histogram).  The
blank-sky distribution is well fit by a Gaussian with mean $\langle
S_{\rm p}\rangle = 0.00 \pm 0.04 {\rm ~mJy~beam}^{-1}$ and rms $\sigma
= 0.48 \pm 0.04 {\rm ~mJy~beam}^{-1}$ (dotted curve) consistent with
NVSS image fluctuations.  The whole distribution of peak flux densities on
the QSO positions is clearly displaced from zero and has a long
positive tail, indicating that the NVSS has detected most $0.2 < z <
0.3$ QSOs statistically at typical peak flux densities $S_{\rm p} \sim
0.3 {\rm ~mJy~beam}^{-1}$, which corresponds to an average spectral
luminosity $\log[L_{\rm 1.4\,GHz}({\rm W\,Hz}^{-1})] \sim 22.7$ in
this redshift range.  

Our high detection rate is not consistent with a power-law
extrapolation of the flat flux-density distribution of AGN-dominated
QSOs, which would yield the dashed curve in Figure~\ref{s2nfig}. The
high detection rate of faint sources requires a peak or ``bump'' in
the luminosity function near $\log[L_{\rm 1.4\,GHz}({\rm W\,Hz}^{-1})]
\sim 22.7$.  This bump is confirmed by individual detections of
nearly all of the $0.2 < z < 0.3$ QSOs at 6~GHz \citep{kim11}.

\begin{figure}
\includegraphics[width=\linewidth,trim=30 40 110 120,clip ]{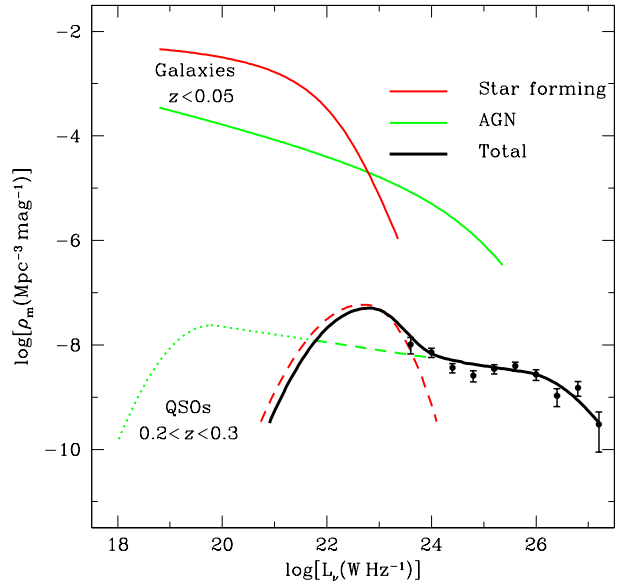}
\caption{Comparison of the 1.4 GHz luminosity functions of nearby
  galaxies whose radio sources are powered primarily by star formation
  (continous red curve) or by AGNs (continuous green curve) with our model for
  the luminosity function of $0.2 < z < 0.3$ QSOs (black curve).   The data
  points are from Figure~\ref{lumffig} with $\log[\rho_{\rm m}({\rm
      Mpc}^{-3}~{\rm mag}^{-1})$ lowered by 0.20 to correct for
    density evolution (Sec.~\ref{lumfsubsec}).  The green dashed line
    is the power-law extrapolation of the AGN-only contribution, and
    the green dotted curve shows how this extrapolation might be
    truncated so the number of AGNs does not exceed the total number
    of QSOs.  The red dashed curve shows the luminosity function of
    the starbursts only.  The black curve is the convolution of the
    AGN and starburst curves. \label{rhomfig}}
\end{figure}

We hypothesize that the bump in the QSO luminosity function
corresponds to the emergence of the host starburst galaxies as the
dominant contributors to the radio emission.  Figure~\ref{rhomfig}
compares the 1.4 GHz luminosity functions of low-redshift ($z < 0.05$)
UGC galaxies \citep{con02} whose radio emission is powered primarily
by AGNs (continuous green curve) or by recent star formation
(continuos red curve) with a simple model for the radio luminosity
function of $0.2 < z < 0.3$ QSOs.  The continuous black curves in
Figures~\ref{pofdfig} and \ref{rhomfig} correspond to our model
luminosity function that fits both the NVSS detections and the $S_{\rm
  p}$ distribution of nondetections.  Our model assumes that the total
radio emission of each QSO is the sum of statistically independent
AGN- and starburst-powered components.  The lower green curve in
Figure~\ref{rhomfig} is similar to the extrapolated luminosity
function of Figure~\ref{lumffig}, constrained by NVSS detections above
$\log[L({\rm W\,Hz}^{-1})] \approx 24$, constrained statistically in
the luminosity range $22 < \log[L({\rm W\,Hz}^{-1})] < 24$ (dashed
green line), and at lower luminosities by requiring that the number of
radio sources equal the number of QSOs (dotted green curve). The
dashed red curve is the radio luminosity function of sources powered
by recent star formation in the host galaxies needed to match the
observed total luminosity function.  The direct detections of faint
sources reported in \citet{kim11} match the dashed red curve very
well.  It suggests that most QSOs host starbursts whose luminosities
are comparable with the more luminous starbursts in nearby galaxies.
The typical star-formation rate can be estimated from the peak of the
dashed red curve: $\log[L_{\rm 1.4~GHz}({\rm W\,Hz})] \sim 22.7$
corresponds to a star-formation rate $SFR \sim 20 M_\odot
{\rm\,yr}^{-1}$ \citep{con92}.

\subsection{Stacking}\label{stackingsec}

We note that measuring the full flux-density distribution of undetected
QSOs is not the same as ``stacking,'' which collapses the distribution
to a single statistic, the mean value of the peak flux density
$\langle S_{\rm p} \rangle$ of the $N$ QSOs fainter than the catalog
limit $S_{\rm lim}$.  The implicit assumptions needed to justify the use of
stacking are (1) $\langle S_{\rm p} \rangle$ is a good statistic for
describing the QSO flux-density distribution and (2) stacking divides
the error in this mean by about $N^{1/2}$.  These assumptions fail if
the flux-density distribution is a nearly flat power law, as we found
for the cataloged NVSS QSOs. When $n_{10}(S)$ is nearly independent of $S$,
then $n(S) \sim S^{-1}$ and $\langle
S_{\rm p} \rangle$ is dominated by the small number $\eta \ll N$ of
sources just fainter than $S_{\rm lim}$, the fractional statistical
uncertainty in $\langle S_{\rm p} \rangle$ is about $\langle S_{\rm p}
\rangle / \eta^{1/2} \gg \langle S_{\rm p} \rangle /N^{1/2}$, and
$\langle S_{\rm p} \rangle$ depends more sensitively on the survey
limit $S_{\rm lim}$ than on the intrinsic properties of the optically
selected QSO sample.

Consider a QSO sample whose differential flux-density distribution is
close to $n(S) \propto S^{-1}$ over a wide ($S_{\rm min} \ll S_{\rm
  max}$) flux-density range $S_{\rm min}$ to $S_{\rm max}$.  The value
of $S_{\rm min}$ is too low to measure; it is only inferred from
the fraction $f$ of detected QSOs and the integral constraint that all
QSOs must be stronger than $S_{\rm min}$.  The average flux density 
$\langle S \rangle$ of
unresolved sources (so $S_{\rm p} \approx S$) in a stack of images with
$S < S_{\rm lim}$ is
\begin{equation}
\langle S \rangle = {\int_{S_{\rm min}}^{S_{\rm lim}} S n(S) dS \over
\int_{S_{\rm min}}^{S_{\rm lim}} n(S) dS} \approx {S_{\rm lim} - S_{\rm min} 
\over \ln(S_{\rm lim} / S_{\rm min})}
\end{equation}
If the fraction $f$ of individually detected sources is not close to
unity, then $S_{\rm lim} \gg S_{\rm min}$ and the
numerator $S_{\rm lim} - S_{\rm min} \approx S_{\rm lim}$.  
The number of sources per logarithmic flux-density
interval is constant between $S_{\rm min}$ and $S_{\rm max}$ when
$n(S) \propto S^{-1}$ so
\begin{equation}
 \ln\biggl({S_{\rm max} \over S_{\rm lim}}\biggr) = f
\ln\biggl({S_{\rm max} \over S_{\rm min}}\biggr)~.
\end{equation}
Using
\begin{equation} \ln \biggl({S_{\rm max} \over S_{\rm min}}\biggr) = 
\ln \biggl({S_{\rm max} \over S_{\rm lim}}\biggr) +
\ln \biggl({S_{\rm lim} \over S_{\rm min}}\biggr)
\end{equation}
yields
\begin{equation}\label{stackfluxeqn}
\langle S_{\rm p} \rangle \approx {S_{\rm lim} \over (1-f) \ln 
(S_{\rm max} / S_{\rm min})}~.
\end{equation}
The stacking flux density $\langle S_{\rm p} \rangle$ depends almost
linearly on the detection limit $S_{\rm lim}$ and only logarithmically
on the intrinsic QSO flux-density distribution.  For example, the NVSS
detected $f = 37/179 \approx 0.21$ of the DR7 QSOs with $0.2 < z <
0.3$ above $S_{\rm lim} = 2.4$~mJy.  If $S_{\rm max} / S_{\rm min}
\sim 10^8$, then Equation~\ref{stackfluxeqn} predicts $\langle S_{\rm
  p} \rangle \approx S_{\rm lim} / 15 \approx 0.16 {\rm ~mJy~beam}^{-1}$,
half of which is contributed by the small expected number $\eta
\approx 5$ of sources with $S_{\rm lim} / 2 < S < S_{\rm lim}$. 

\citet{whi07} made numerical simulations of stacking with Gaussian and
exponential distributions and also concluded that the stack mean does
not provide a very robust measurement.  They studied the stack median
as well, which should be better than the mean because it is less
sensitive to the strongest sources.  However, their simulations showed
that the stack median recovered from exploiting low-SNR data (the
usual reason for stacking) approaches the true mean, so they
interpreted their stack medians as means.

\citet{mah12} has independently noted several limitations of stacking.

We believe it is both safer and more sensitive to fit a model to the
entire amplitude distribution (Figure~\ref{pofdfig}) than to rely on
{\it any} stacking statistic.  In particular, if the whole $S_{\rm p}$
distribution is shifted to positive flux densities, we can conclude
that {\it most} of the QSOs are detectable radio sources.

\subsection{Evolution of the Faint Radio Sources, $0.2 < z < 0.45$}

\begin{figure}
\includegraphics[width=4.0in,trim=110 40 140 50,clip]{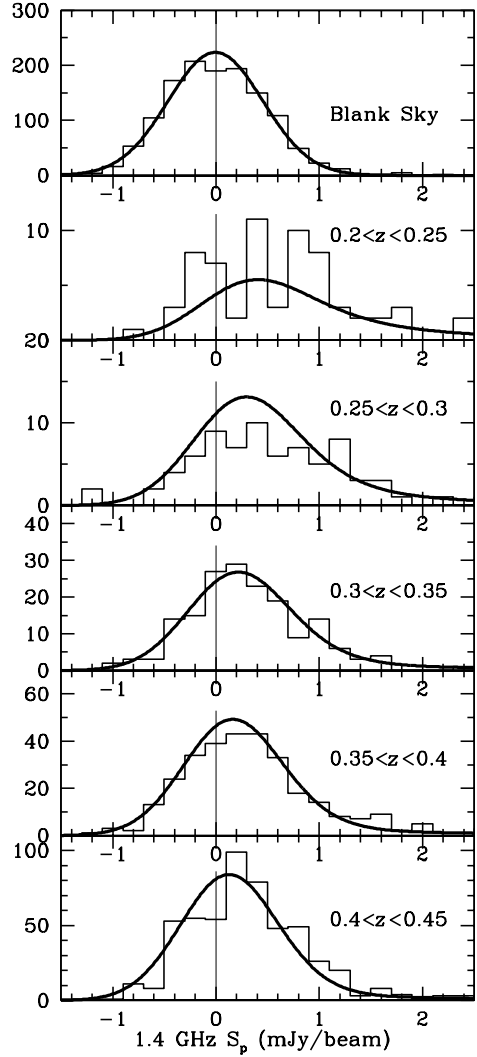}
\caption{Histograms of NVSS peak flux densities for 1313 SDSS QSOs
  in five redshift ranges spanning $0.2 < z < 0.45$ fitted with an evolutionary
model (smooth curves).  The top panel
  shows the distribution of peak flux densities on ``blank sky''
  reference positions one degree north of the QSOs; it is nearly
  Gaussian with mean $\langle S_{\rm p}\rangle = -0.004 \pm 0.013 {\rm
    ~mJy~beam}^{-1}$ and rms $\sigma = 0.458 \pm 0.013 {\rm
    ~mJy~beam}^{-1}$.  \label{pofdevolfig}}
\end{figure}

Encouraged by our statistical detection of 1.4 GHz emission at
starburst luminosities from most of the 179 QSOs with $0.2 < z < 0.3$
and its confirmation by individual detections \citep{kim11},
we measured NVSS peak flux densities at the positions of all 1313 QSOs
with $0.2 < z < 0.45$ and divided them into five redshift bins ($0.2 <
z < 0.25$, $0.25 < z < 0.3$, $0.3 < z < 0.35$, $0.35 < z < 0.4$, and
$0.45 < z < 0.5$) to constrain the evolution of the hypothesized
starburst radio emission over lookback times from 2.4 to 4.6 Gyr.  The
results are shown as histograms in Figure~\ref{pofdevolfig}.  The
histogram of peak flux densities in 1313 ``blank sky'' positions one
degree north of the QSOs is plotted in the top panel, and the matching
smooth curve shows the Gaussian fit with mean $\langle S_{\rm
  p}\rangle = -0.004 \pm 0.013 {\rm ~mJy~beam}^{-1}$ and rms $\sigma =
0.458 \pm 0.013 {\rm ~mJy~beam}^{-1}$.  The lower five panels contain
the histograms of NVSS peak flux densities at the QSO positions,
separated into redshift bins.

The NVSS detection rate in the $0.2 < z < 0.45$ QSO sample is only
$163/1313 \approx 12$\% above 2.4 mJy, so 88\% of the 1313 QSOs have
$S_{\rm p} < 2.4$~mJy~beam$^{-1}$ and models that match the numbers of
sources in each redshift bin are constrained primarily by the numbers
and hence density evolution of the optically selected QSOs.  The radio
data determine the flux-density distribution within each redshift bin
and hence constrain the luminosity evolution of these low-luminosity radio
sources.

As in Section~\ref{lumfsubsec}, we describe the density evolution of
the optically selected QSOs by $\rho_{\rm m} (z) \propto
(1+z)^\delta$.  The best fit, shown by the smooth curves in
Figure~\ref{pofdevolfig}, is for $\delta = +8.5 \pm 2$.  The best fit
is poor [$P(\chi^2) \sim 0.01$] because the numbers of optically
selected QSOs in the redshift bins $0.2 < z < 0.25$ ($N=85$) and $0.25
< z < 0.3$ ($N=94$) are nearly equal but the comoving volumes
(0.402~Gpc$^3$ and 0.572~Gpc$^3$, respectively) are not.  These
numbers favor slightly negative density evolution ($\delta < 0$) in
the $0.2 < z < 0.3$ redshift range.  The positive value $\delta = +8.5
\pm 2$ for the faint sources over the whole range $0.2 < z < 0.45$ is
consistent with the value $\delta = +6.0 \pm 1.8$ for the stronger
(primarily AGN-powered) radio sources with $0.2 < z < 0.45$.

If we model 1.4 GHz luminosity evolution of the fainter radio sources
by the form $L_\nu (z) \propto (1+z)^\lambda$, the best fit is
consistent with no luminosity evolution: $\lambda = -1 \pm 3$.

All of these results are actually consistent with each other and can
be summarized as follows: The optically selected QSOs undergo strong
density evolution over the reshift range $0.2 < z < 0.45$.  Both the
weak and strong radio sources undergo a comparably strong density
evolution with no detectable luminosity evolution.  The likelihood of
radio emission at any spectral luminosity from any particular QSO is
independent of redshift in the range $0.2 < z < 0.45$. Although the
space density of optically selected QSOs evolves dramatically, the
likelihood and strength of radio emission from each QSO appears
to be independent of redshift.

\section{The Magnitude-limited $1.8 < z < 2.5$ QSO sample}\label{hizsec}

Many historical QSO samples favor the redshift range $1.8 < z < 2.5$
in which the Ly$\alpha$ line is redshifted to the blue optical band.
To test the statistical significance of the \citet{mil90} bimodal radio
luminosity distribution of such high-redshift QSOs, we chose a
magnitude-limited sample of all 2471 color-selected DR7 QSOs brighter
than $m_{\rm r} = 18.5$ in that redshift range.  These QSOs are all
extremely luminous because our red magnitude limit corresponds to $M_{\rm i}
\approx -26.9$ at $z = 1.8$ and $M_{\rm i} \approx -27.7$ at $z =
2.5$.

\begin{figure}
\includegraphics[width=\linewidth,trim=28 10 60 50,clip ]{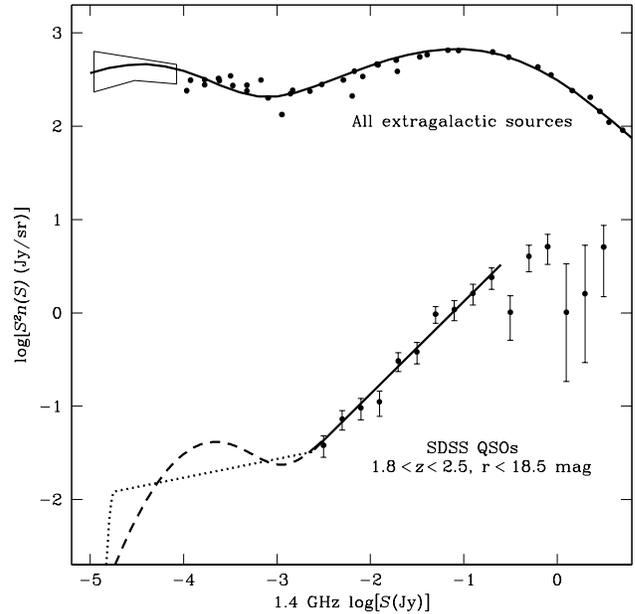}
\caption{The peaks in the brightness-weighted count of all
  extragalactic sources near $\log[S({\rm Jy})] = -1$ and $\log[S({\rm
      Jy})] = -4$ match the expected contributions from AGNs
  (primarily radio galaxies) and star-forming galaxies, respectively,
  at typical redshifts $z \sim 1$.  The high-redshift SDSS QSO count
  (filled points with error bars in the lower part of the plot) of
  sources stronger than the NVSS detection limit $S = 2.4{\rm ~mJy}$
  is well fit by a power-law (solid line) for which the slope of $S^2
  n(S)$ is $+0.99 \pm 0.02$.  The dashed and dotted curves illustrate
  the range of faint-source counts needed to match the NVSS peak
  flux-density distribution on the positions of high-redshift QSOs
  fainter than $2.4{\rm ~mJy}$.  Abscissa: log 1.4 GHz flux density
  (Jy).  Ordinate: Differential source count multiplied by
  $S^2$ (Jy~sr$^{-1}$). \label{s2nhizfig}}
\end{figure}

The NVSS detected radio emission stronger than $S = 2.4{\rm ~mJy}$
from only 191 (8\%) of them (Table \ref{nvsshiztable}), consistent
with the usual low detection rates at this sensitivity.  The
brightness-weighted 1.4 GHz flux-density distribution $S^2 n(S)$ of
the NVSS detections is shown by the points with error bars in the
lower part of Figure~\ref{s2nhizfig}.  The power-law fit to their
brightness-weighted differential source count is
\begin{equation}\label{hizcounteq}
S^2 n(S) = (12.9 \pm 1.0) \times S^{0.99
  \pm 0.02} {\rm ~Jy~sr}^{-1}
\end{equation}
in the flux-density range $2.4 < S{\rm (mJy)} < 1000$.  Once again,
the QSO counts are extremely flat in the sense that the number of
sources per decade of flux density is nearly constant: $S n(S) \propto
n_{10}(S) \propto S^{-0.01}$.  The strongest 13 of the 2471 QSOs
($\approx 0.5$\%) account for more than half of the total radio flux
density from the high-redshift QSO sample.

\begin{figure}
\includegraphics[width=\linewidth,trim=40 30 110 100,clip ]{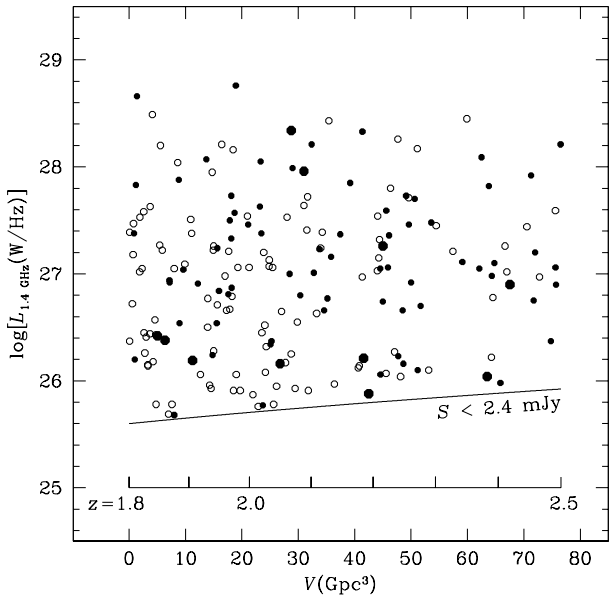}
\caption{The 1.4 GHz luminosity distribution of QSOs detected by the
NVSS is shown as a function of the volume enclosed between the
redshift of each QSO and the sample's lower redshift limit $z = 1.8$.
The $S = 2.4 {\rm ~mJy}$ NVSS sensitivity limit is indicated by the
curve.  Symbols distinguish QSOs by their absolute magnitudes: large
filled symbols for $M_{\rm i} < -28.7$, small filled symbols for
$-28.7 < M_{\rm i} < -27.7$, and open symbols for $-27.7 < M_{\rm
  i}$. The QSO sample is volume limited for QSOs with $M_{\rm i} <
-27.7$ (filled points), so the density of filled points above the curve
is proportional to the 1.4 GHz luminosity function of QSOs more
luminous than $M_{\rm i} = -27.7$. 
Abscissa: comoving volume $V$ (Gpc$^3$).  Ordinate: log 1.4 GHz
spectral luminosity.  Parameter: redshift $z$.
\label{lovhizfig}}
\end{figure}

\begin{figure}
\includegraphics[width=\linewidth,trim=20 30 80 100,clip ]{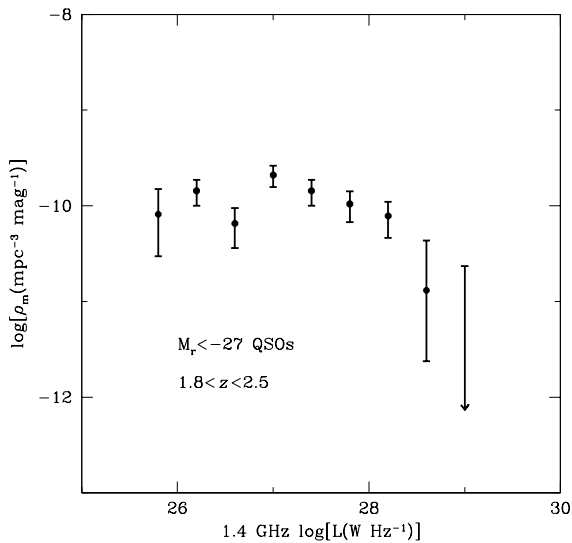}
\caption{The 1.4 GHz luminosity function of QSOs more luminous
than $M_{\rm i} = -27.7$ in the redshift
range $1.8 < z < 2.5$ is shown by data
points at $\log[L_{\rm 1.4\,GHz}({\rm W\,Hz}^{-1})] = 25.8$ to 28.6 by 0.4 
and an upper limit at 29.0.  
Abscissa: log 1.4 GHz spectral luminosity (W Hz)$^{-1}$.  Ordinate:
log comoving space density (Mpc$^{-3}$) of sources per ``magnitude'' = dex(0.4)
luminosity range.
  \label{lumfhizfig}}
\end{figure}

Figure~\ref{lovhizfig} shows the luminosity distribution of these QSOs
as a function of the comoving volume beyond $z = 1.8$ in $\Omega =
2.66{\rm ~sr}$.  Our magnitude-limited high-redshift QSO sample is
volume limited for those QSOs brighter than $M_{\rm i} = -27.7$
(filled points) above the $S = 2.4{\rm ~mJy}$ line, where the plotted
density of filled points is proportional to the evolving radio
luminosity function.  In the redshift range $1.8 < z < 2.5$ the
density evolution of radio sources stronger than $\log[L({\rm
    W\,Hz}^{-1})] = 26$ in high-redshift QSOs with $M_{\rm i} \leq
-27.7$ is small or slightly negative, as expected: if $\rho_{\rm m} \propto
(1+z)^\delta$, $\delta = -0.3 \pm 1.8$. The correlation of radio and
optical luminosities is weak; if anything, the most luminous radio
sources are under-represented among the optically most luminous QSOs
(large filled points), many of which have radio luminosities near
$\log[L({\rm W\,Hz}^{-1})] = 26$.  As expected from the 1.4 GHz source
count, the 1.4 GHz spectral luminosity function averaged over the $1.8
< z < 2.5$ redshift range is nearly flat (Figure~\ref{lumfhizfig})
until it cuts off around $\log[L_\nu({\rm W\,Hz}^{-1})] \approx 28.8$,
the highest radio luminosity found in this huge (77 Gpc$^3$) comoving
volume.

\subsection{NVSS Statistical Detections: Faint Sources Powered Primarily by Star Formation?}\label{hizpofdsubsec}

\begin{figure}
\includegraphics[width=\linewidth,trim=30 30 30 80,clip ]{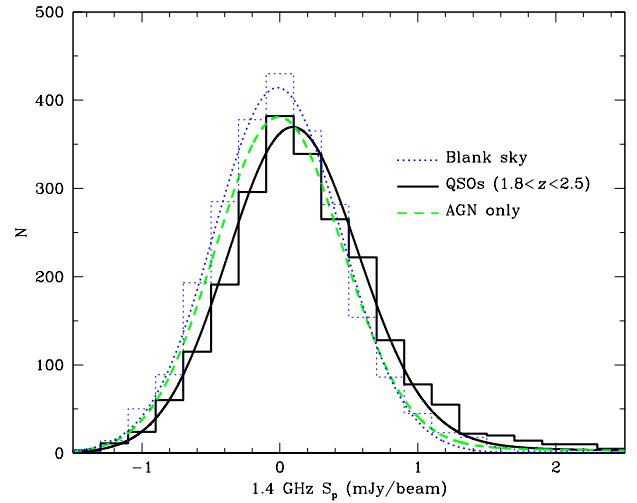}
\caption{The continuous histogram shows the distribution of NVSS peak
  flux densities for the magnitude-limited sample of 2471 SDSS QSOs
  brighter than $m_{\rm r} = +18.5$ in the redshift range $1.8 < z <
  2.5$.  The dotted histogram (colored blue in the online version)
  indicates the distribution of NVSS peak flux densities at ``blank
  sky'' positions one degree north of the QSOs, and it is well fit by
  a Gaussian of mean $\langle S_{\rm p}\rangle = -0.014 \pm 0.009 {\rm
    ~mJy~beam}^{-1}$ and rms $\sigma =0.459 \pm 0.009 {\rm
    ~mJy~beam}^{-1}$ (dotted curve, colored blue in the online
  version). The dashed curve (colored green in the online version) is
  the distribution predicted by extrapolating the flux-density
  distribution of NVSS sources powered primarily by AGNs to lower
  luminosities, and the continuous black curve represents the distribution
  predicted by our best-fit model extremely luminous star-forming host
  galaxies may contribute significantly to the radio emission of
  high-redshift QSOs. \label{hizpofdfig}}
\end{figure}

Using the technique described in Sec.~\ref{pofdsubsec}, we obtained
the NVSS peak flux densities at the positions of all 2471
high-redshift QSOs and at the 2471 ``blank sky'' positions offset by 1
deg of declination.  Figure~\ref{hizpofdfig} shows the observed
distributions of peak flux densities on the QSO positions (continuous
histogram) and on the blank sky points (dotted histogram).  The
continuous histogram is clearly shifted to the right, indicating that
most of the QSOs are weak radio emitters, and their median peak flux
density is $\langle S \rangle \approx 0.05 \pm 0.01\,{\rm mJy\,beam}^{-1}$.  The
blank sky distribution is well approximated by a Gaussian (dotted
curve) with mean $\langle S_{\rm p} \rangle = -0.014 \pm 0.009{\rm
  ~mJy~beam}^{-1}$ and rms $\sigma = 0.459 \pm 0.009{\rm
  ~mJy~beam}^{-1}$.  Simply extrapolating the flat power-law
flux-density distribution (Equation~\ref{hizcounteq}) of sources so
luminous \{$\log[L({\rm ~W~Hz}^{-1})] > 25.8$\} that they must be
powered primarily by AGNs and convolving that flux-density
distribution with the blank-sky Gaussian fit yields the dashed curve
in Figure~\ref{hizpofdfig}.  The dashed curve is a poor match to the
continuous QSO histogram because the extrapolation predicts that the
peak flux densities of most QSOs are extremely low ($< 0.001 {\rm
  ~mJy~beam}^{-1}$).  It is necessary to invoke a rise or peak in the
flux-density distribution near $S \sim 0.05$~mJy ($R \sim 0.3$) 
to explain the high
detection rate implied by the positive offset of the QSO histogram.

The dotted line in Figure~\ref{s2nhizfig} indicates the weighted
faint-source count from the power-law extrapolation that (1) matches
with the observed count at 2.4 mJy, (2) satisfies the integral
constraint that the total number of radio sources must equal the total
number of QSOs, and (3) best fits the observed flux-density
distribution of faint sources; it is
\begin{equation}
n(S) = 0.11 S^{-1.8} {\rm ~Jy}^{-1}{\rm ~sr}^{-1} , S > S_{\rm min}
\end{equation}
and $n(S) = 0$ below $S_{\rm min} = 0.018 {\rm ~mJy}$.  Of course,
there is no reason to expect a break in the count slope
to occur at the NVSS catalog limit, $2.4{\rm ~mJy}$.  At the midpoint
of the QSO redshift range, $z = 2.15$, that break corresponds to a
spectral luminosity $[\log L ({\rm W~Hz}^{-1})] \approx 25.8$, which
is higher than the spectral luminosities of even the strongest
starbursts.  

Any more realistic extrapolation with a continuous count and count
slope at 2.4 mJy must have a peak in $S^2 n(S)$ at lower flux
densities to match the statistical data, as shown by an illustrative
model that yields the dashed curve in Figure~\ref{s2nhizfig}.
Convolving the model counts with the NVSS image noise distribution
yields a good fit to the observed distribution of QSO peak flux
densities; it is shown by the continuous curve in
Figure~\ref{hizpofdfig}).  The median flux density of our
high-redshift QSOs in this model is $\langle S \rangle \approx
0.05{\rm ~mJy}$, so the median 1.4 GHz spectral luminosity 
in the source frame (calculated assuming a median spectral index
$\alpha = -0.7$)
is $\log[(L_\nu({\rm W~Hz}^{-1})] \approx 24.1$ at $z =
2.15$.  This is a factor of four higher than the 1.4 GHz spectral
luminosity of Arp 220, and it might come from an AGN or a starburst.
If it originates in a starburst obeying the FIR/radio correlation, an
ultraluminous [$(L / L_\sun) \sim 4 \times 10^{12}$] but not
hyperluminous [$(L / L_\sun) > 10^{13}$] starburst is required.

\section{Is QSO Radio Emission Bimodal?}

\begin{figure}
\includegraphics[width=\linewidth,trim=0 40 60 110,clip ]{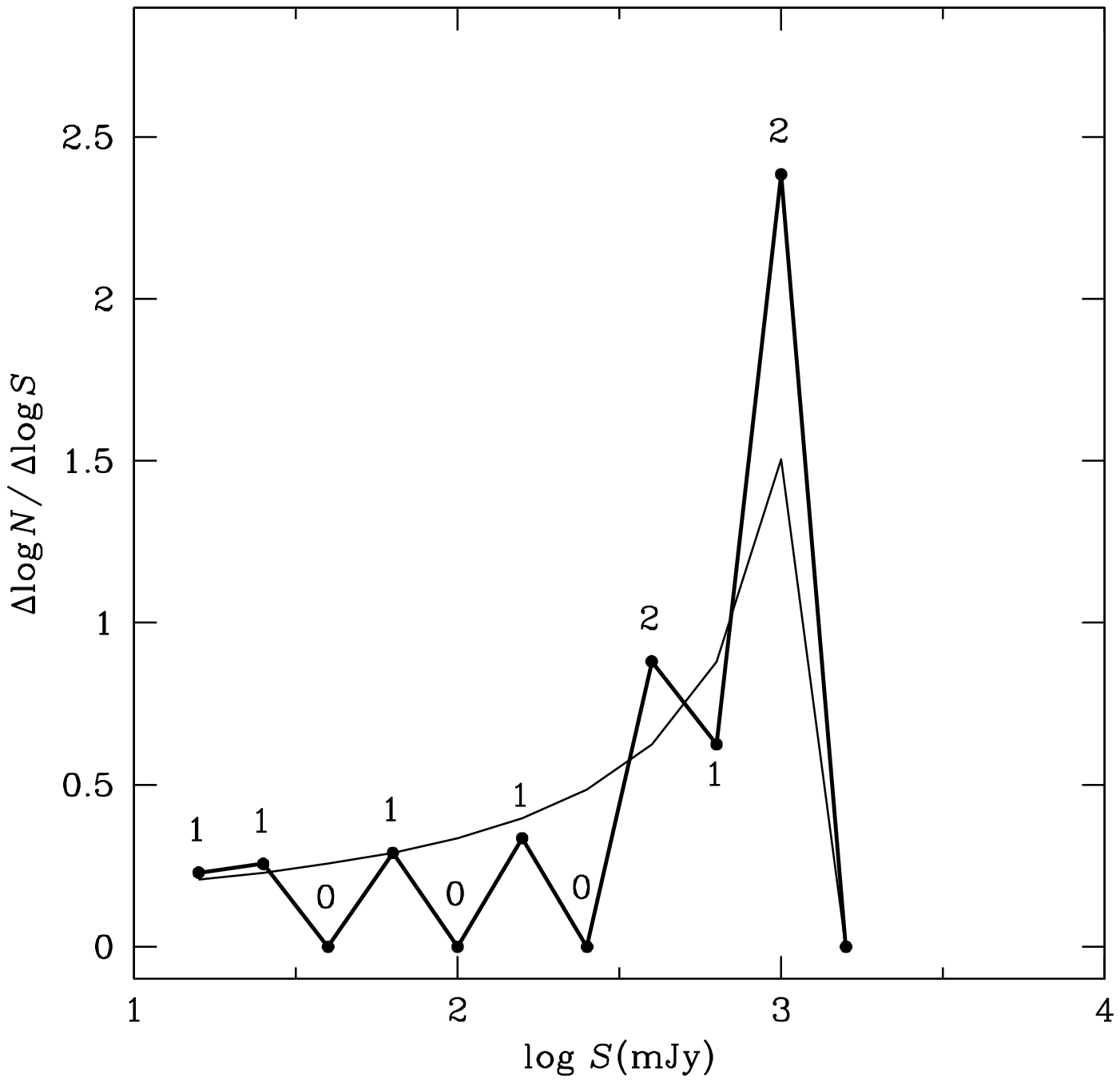}
\caption{The differential flux-density distributions of QSOs as plotted
by \citet{str80} with numbers of QSOs per bin added.
\label{strittmatterfig}}
\end{figure}

The first published claim of a bimodal flux-density distribution \citep{str80}
features a strong-source peak too narrow to be consistent with
relativistic beaming.  The \citet{str80} flux-density distribution
plot is reproduced in Figure~\ref{strittmatterfig}.  The narrow peak
in the quantity $\Delta \log N / \Delta \log S$ (where $N$ is the
number of sources stronger than $S$ and $\Delta \log S = 0.2$ is the
logarithmic width of each flux-density bin) near $\log[S({\rm mJy})] =
3$ is an apparently significant factor of dex$(2) \sim 100$ higher
than the long tail of fainter sources. We have inserted the actual
numbers of radio sources contributing to each logarithmic flux-density
bin into Figure~\ref{strittmatterfig}. The peak is produced by only
two sources, as $N$ goes from 1 to 3, so $\Delta \log N =
\log(3)-\log(1) \approx 0.48$ and $\Delta \log N / \Delta \log S
\approx 0.48 / 0.2 \approx 2.4$ as plotted.  However, the quantity
$\Delta \log N / \Delta \log S$ is not the usual logarithmic
differential source count or flux-density distribution $\log \Delta N
/ \Delta \log S$. The thin curve shows the distribution of $\Delta
\log N / \Delta \log S$ that would result from the same total number of
radio sources (ten) having a constant $\log \Delta N / \Delta \log S$
in the plotted flux-density range.  Thus there is no significant peak
in the nearly constant logarithmic differential flux-density
distribution of these QSOs, which is consistent with our
Equation~\ref{hizcounteq}.

\citet{mil90} made sensitive 5 GHz observations of 105 bright QSOs
with $1.8 < z < 2.5$ selected by their emission lines.  They found
nine with $\log[L({\rm W\,Hz}^{-1})] > 26$ but none with $25 <
\log[L({\rm W\,Hz}^{-1})] < 26$.  All nine detections have $S > 20{\rm
  ~mJy}$, far above their $3 \sigma \sim 0.6{\rm ~mJy}$ detection
limit.  This is evidence for bimodal flux-density and luminosity
distributions, but at what statistical significance?  The nine
detections have flux densities 21, 28, 28, 51, 76, 98, 105, 111, and
136 mJy.  The likelihood of no weaker detections above 0.6 mJy depends
on how the flux densities are binned.  The bin limits should not be
chosen {\it a posteriori} to minimize this probability.  In bins of
width dex(1) starting at the detection limit, there are $N = 0$
detections with $0.6 < S({\rm mJy}) < 6$, $N = 4$ detections with $6 <
S({\rm mJy}) < 60$, and $N = 5$ detections with $60 < S({\rm mJy}) <
600$.  If the flux-density distribution of $1.8 < z < 2.5$ QSOs is
flat in logarithmic bins (as in Equation~\ref{hizcounteq}), there should be
equal numbers in each bin.  The average number of detections per bin
is $\langle N \rangle = 3$, so the Poisson probability of no
detections in the faintest bin is $P_{\rm P} (0,3) = \exp(-3) \approx
0.05$.  Thus the lack of detections fainter than 6 mJy in this small
sample may just be the result of bad luck, which we did not encounter
in our much larger high-redshift QSO sample.

We prefer to study evolving QSO luminosity functions, which are
intrinsic properties of QSOs, rather than flux-density distributions
or luminosity distributions, which depend on observational parameters.
We found that the luminosity functions $\rho_{\rm m}(L)$ of radio sources
so powerful that they must be produced by AGN are smooth and nearly
flat power laws with no peaks or other features.  There are ``bumps''
in the radio luminosity functions of QSOs, but they appear only at such
low radio luminosities that the resulting radio flux densities are well
below 1 mJy.  The ``bump'' luminosities are consistent with
significant starburst contributions
(recall Figure~\ref{rhomfig}): $\log[L({\rm W~Hz}^{-1})] \sim
22.7$ for our moderate-luminosity QSOs at low ($0.2 < z < 0.45$)
redshifts and $\log[L({\rm W~Hz}^{-1})] \sim 24.1$, corresponding
to an $(L/L_\odot) \sim 4 \times 10^{12}$ ultraluminous starburst, for the most
luminous QSOs at higher redshifts, $1.8 < z < 2.5$.

\section{The AGN and Host Galaxy Components of QSO Radio 
Emission}\label{qsolumfsec}

The radio emission from a QSO is the sum of contributions from its AGN
and from star formation in its host galaxy.  If the luminosities of
the two contributions are nearly independent of each other at every
redshift, the spectral luminosity function of QSOs will be the
convolution of the AGN and starburst spectral luminosity functions.

\subsection{Low-redshift QSOs}

The AGN-powered portion of the luminosity function in the redshift
range $0.2 < z < 0.45$ (Figure~\ref{lumffig}) was estimated from 163
NVSS detections of luminous sources.  There are only 37 NVSS
detections in the narrow redshift range $0.2 < z < 0.3$, too few for
statistical accuracy, so we estimated the $0.2 < z < 0.3$ AGN
luminosity function from the evolving AGN luminosity function in the
redshift range $0.2 < z < 0.45$. The luminosity function in the range
$0.2 < z < 0.3$ is lower than the $0.2 < z < 0.45$ luminosity function
by $\Delta \log \rho_{\rm} = -0.20$ (Sec.~\ref{lumfsubsec}). This
estimate is shown by the data points with error bars in
Figure~\ref{rhomfig}.  For comparison, the 1.4 GHz local ($z \lesssim
0.05$) luminosity functions of galaxies \citep{con02} whose radio
emission is dominated either by AGNs (continuous green curve) or by
star formation (continuous red curve) are also shown. The radio
luminosity function produced by AGNs in local galaxies is flatter and
extends to higher radio luminosities than the luminosity function of
star-forming galaxies, so local AGN-powered radio sources are more
common above $\log[L_{\rm 1.4~GHz}({\rm W\,Hz}^{-1})] = 23$ and
star-forming galaxies have higher space densities than AGNs at lower
luminosities.  The luminosity $\log[L_{\rm 1.4~GHz}({\rm W\,Hz}^{-1})]
\approx 22$ of the galaxy M82 is typical for a nearby radio-selected
star-forming galaxy.

Even allowing for starburst luminosity evolution, it is likely that
the radio emission of low redshift ($0.2 < z < 0.45)$ QSOs stronger
than $\log[L_{\rm 1.4~GHz}({\rm W\,Hz}^{-1})] = 24$ is dominated by
AGNs.  To estimate the luminosity function that would result if all
QSO radio emission were powered by only by AGNs, even at lower
luminosites, we made the smooth power-law extrapolation (dashed green
line in Figure~\ref{rhomfig}) of the observed QSO luminosity function to
lower luminosities.  This extrapolation must be truncated below
$\log[L_{\rm 1.4~GHz}({\rm W\,Hz}^{-1})] \sim 20$, as suggested by the
dotted green curve in Figure~\ref{rhomfig}, to keep the number of
radio sources from exceeding the total number of low-redshift QSOs
(1313).  The details of the AGN luminosity function below $\log[L_{\rm
    1.4~GHz}({\rm W\,Hz}^{-1})] \approx 22$ are not well known but
don't matter because any luminosity function in this range always
predicts that most of the low-redshift QSOs powered by entirely by AGN
would be too radio quiet ($< 0.02 {\rm ~mJy}$) to be statistically
detectable by the NVSS.  The dashed curve in Figure~\ref{pofdfig}
shows the AGN-only flux-density distribution predicted by this model
for the $0.2 < z < 0.3$ QSOs.  The distribution peak flux density $\langle
S_{\rm p}\rangle \approx 0.05 {\rm ~mJy~beam}^{-1}$ is much lower than
the observed $\langle S_{\rm p}\rangle \approx 0.33 {\rm
  ~mJy~beam}^{-1}$.  The AGN-only distribution is distinguishable from
the blank-sky distribution (dotted curve) only because its height has
been depressed by the loss of QSOs in the long tail of strong sources
above $2.5 {\rm ~mJy~beam}^{-1}$, which are off scale on this plot.
The failure of the dashed curve to fit the data suggests that most of
the low-redshift QSOs are not in ``red and dead'' elliptical galaxies
with little or no ongoing star formation.

{\it Any} QSO luminosity function that is consistent with both the
NVSS direct detections of sources stronger than 2.4~mJy and the
distribution of peak flux densities in Figure~\ref{pofdfig} must rise
sharply just below $\log[L_\nu({\rm W\,Hz}^{-1})] = 24$ and fall
fairly quickly at lower luminosities lest it imply more than 1313 SDSS
QSOs with $0.2 < z < 0.45$.  \citet{kim11} proposed that radio
emission from the star-forming host galaxies is the cause of that
rise.  The peak of the the QSO host-galaxy radio luminosity function
(dashed red curve in Figure~\ref{rhomfig}) is at $\log[L_\nu({\rm
    W\,Hz}^{-1})] \approx 22.7$.  If the faint radio sources are
dominated by emission from the star-forming host galaxies, their radio
spectral indices between 1.4~GHz and 6~GHz should be $\alpha \approx
-0.7$, which is the observed value \citep{kim11}.

As it must, the radio luminosity function of the low-redshift QSO host
galaxies lies below the radio luminosity function of {\it all} nearby
galaxies powered primarily by star formation, but it is not much lower
at the highest radio luminosities because many low-redshift QSOs seem
to have high star-formation rates.  The radio luminosity function of
QSO host galaxies must fall rapidly below $\log[L_{\rm 1.4~GHz}{\rm
    W~Hz}^{-1} \sim 20$ to satisfy the constraint that its integral
  not exceed the total number of SDSS QSOs in $0.2 < z < 0.3$.  In
  this respect, it differs from the radio luminosity function of all
  nearby star-forming galaxies (red curve in Figure~\ref{rhomfig}),
  which rises slowly but monotonically even at very low luminosities.
  This difference is not surprising: most of the faintest nearby radio
  sources lie in low-mass galaxies with low star-formation rates
  \citep{con02}, while QSOs are found only in massive galaxies.  The
  real surprise 
to radio astronomers accustomed to comparing QSOs with radio galaxies
is that massive elliptical galaxies with low
  star-formation rates are so rare in our low-redshift QSO sample.

\subsection{High-redshift QSOs}

The high statistical detection rate of $1.8 < z < 2.5$ QSOs reported
in Sec.~\ref{hizpofdsubsec} shows that most of these very luminous
(median $\langle M_{\rm i}\rangle \approx -27.5$) high-redshift QSOs
contain radio sources with median spectral luminosity $\log[L_{\rm 1.4
    ~GHz}({\rm W~Hz}^{-1})] \approx 24.1$.  This high detection rate
is not consistent with a power-law extrapolation from higher radio
luminosities of the AGN luminosity function; there must also be a
``bump'' in the QSO radio luminosity function at high redshifts.
As illustrated in Figure~\ref{s2nhizfig}, the bump must lie in
a fairly narrow flux-density range $-5 < \log[S(Jy)] < -3$ to be consistent
with the distribution of NVSS peak flux densities (Figure~\ref{hizpofdfig}).  
Such a bump is consistent with, but does not by itself require, radio
emission from ultraluminous [$(L/L_\sun) \sim 4 \times 10^{12}$]
starbursts in the host galaxies of the most luminous QSOs.  Should we
expect that most luminous, high-redshift QSOs contain ultraluminous
starbursts?

The presence of a luminous QSO implies a high rate of radiatively
efficient accretion onto a supermassive black hole.  During ``quasar
mode'' or ``cold mode'' accretion \citep{cro06}, cold gas is rapidly
fed via an accretion disk to the black hole.  The reservoir of cold
gas that feeds the AGN may also produce a burst of star formation, as
suggested by recent observations starbursts in the host galaxies of
high-luminosity ($L_{\rm x} > 10^{37}$~W) X-ray QSOs with $z \gtrsim
2$ \citep{rov12} and of infrared emission from cold dust in the host
galaxies of the fairly luminous ($\langle M_{\rm i} \rangle
\approx -25$) QSOs found in the Herschel ATLAS survey \citep{bon11}.
\citet{bon11} associated the total ($8\,\mu < \lambda < 1000\,\mu$)
infrared luminosity $L_{\rm IR}$ with the star-formation luminosity
and found it is related to the QSO luminosity $L_{\rm QSO}$ via
\begin{equation}\label{boneq}
L_{\rm IR} \propto L_{\rm QSO}^\theta (1 + z)^\zeta ~,
\end{equation}
where $\theta = 0.22 \pm 0.08$ and $\zeta = 1.6 \pm 1$.  Using the
FIR/radio correlation \citep{con92} and the median 1.4 GHz spectral
luminosity, we estimate the median starburst luminosity $\log(L_{\rm
  IR}/L_\sun) \approx 11.2$ for our low-redshift QSOs at typical
redshift $z \approx 0.35$.  The median absolute magnitudes of the low-
and high-redshift QSOs are $M_{\rm i} = -23.38 \pm 0.03$ and $M_{\rm
  i} = -27.48 \pm 0.01$, for a QSO luminosity ratio $\approx 44$.
Inserting these values into Equation~\ref{boneq} indicates that our
high-redshift ($z \approx 2.15$) QSOs should have a median starburst
luminosity $\log(L / L_\sun) = 12.2 \pm 0.4$ and a median radio
spectral luminosity $\log[L_{\rm 1.4~GHz}({\rm W~Hz}^{-1})] = 23.7 \pm
0.4$, only one sigma below our observed median $\log[L_{\rm
    1.4~GHz}({\rm W~Hz}^{-1})] = 24.1$ for SDSS QSOs in the redshift
range $1.8 < z < 2.5$.

\section{Summary}

We investigated the 1.4 GHz radio emission from large optically
selected samples of SDSS DR7 QSOs in two redshift ranges: $0.2 < z <
0.45$ and $1.8 < z < 2.5$.  As expected, only about 10\% of the QSOs
are sufficiently radio loud to be detected above the $S = 2.4{\rm
  ~mJy}$ NVSS catalog limit. Such sources are so luminous
that the bulk of their radio emission must be powered by AGNs.  The
1.4~GHz luminosity functions of AGN-powered sources both samples
(Figures~\ref{lumffig} and \ref{lumfhizfig}) are flat power laws, so
flat that the number of sources per decade of flux density or
luminosity is nearly constant.  These luminosity functions show no
features or signs of bimodality that might indicate two or more
distinct QSO types among the $\sim10$\% of QSOs that are in the NVSS
catalog.

Extrapolating the flat luminosity functions to lower luminosities
predicts that most of the undetected QSOs should be extremely faint
radio sources.  However, the distributions of peak flux densities at
QSO positions on the NVSS images reveal that most QSOs are moderately
luminous radio sources, so there must be peaks or ``bumps'' in their
1.4 GHz luminosity functions in both redshift ranges.  QSOs with $0.2
< z < 0.3$ have a median peak flux density $S_{\rm p} \approx 0.3 {\rm
  ~mJy~beam}^{-1}$ (Figure~\ref{pofdfig}) and median spectral luminosity
$\log[L_{\rm 1.4~GHz}({\rm W~Hz}^{-1})] \approx 22.7$; QSOs in the
redshift range $1.8 < z < 2.5$ have a median peak flux density $S_{\rm
  p} \approx 0.05 {\rm ~mJy~beam}^{-1}$ (Figure~\ref{hizpofdfig}) and
median spectral luminosity $\log[L_{\rm 1.4~GHz}({\rm W~Hz}^{-1})]
\approx 24.1$.  The reality of the statistical detections in the
narrow redshift range $0.2 < z < 0.3$ was confirmed by individual
detections of nearly all of these QSOs by the VLA at 6 GHz
\citep{kim11}, and the median spectral index of these sources is
$\langle \alpha \rangle \approx -0.7$, typical of optically thin
synchrotron emission from either AGNs or star-forming galaxies.

We suggest that the faint radio sources found in most bright optically
selected QSOs are primarily powered by star formation.  The median
radio luminosities correspond to star-formation rates $\dot{M} \sim 20
M_\odot {\rm ~yr}^{-1}$ and $\dot{M} \sim 500 M_\odot {\rm ~yr}^{-1}$
for the moderately luminous ($\langle M_{\rm I} \rangle \approx
-23.4$) QSOs with $0.2 < z < 0.45$ and the extremely luminous
($\langle M_{\rm i} \rangle \approx -27.5$) QSOs with $1.8 < z < 2.5$,
respectively.  The total infrared luminosities of such starbursts can
be estimated from the FIR/radio correlation.  They are $\log(L_{\rm
  IR} / L_\sun) \approx 11.2$ for the $0.2 < z < 0.45$ QSOs and
$\log(L_{\rm IR} / L_\sun) \approx 12.6$ for the $1.8 < z < 2.5$ QSOs.
Such powerful starbursts may be fueled by the same cold gas reservoirs
that flow into the central supermassive black holes that power the
QSOs themselves.  Our starburst interpretation of the low-luminosity
QSO radio emission is supported by (1) far-infrared detections of cold
dust in QSO host galaxies \citep{bon11}, and (2) agreement of our
radio results with their empirical scaling of $L_{\rm IR}$ with QSO
luminosity and redshift.

With cm-wavelength radio data alone, it is difficult to test our
hypothesis that star formation in the host galaxies of most QSOs
powers the faint radio sources that make the ``bumps''
in our otherwise flat radio luminosity functions and source counts.
One way would be to show that the faint radio
sources in QSOs follow the tight FIR/radio correlation obeyed by
nearly all starburst galaxies.  However, it is necessary to
distinguish the FIR emission from a large mass of cool dust heated by
stars from the mid-infrared (MIR) emission from a small mass of warmer
dust heated by an AGN \citep{mor10}, so it is preferable that the the
dust flux density be measured in the Rayleigh-Jeans tail at submillimeter
wavelengths, by ALMA for example.  
Alternatively, ALMA submillimeter
spectroscopy of molecular and ionic lines sensitive to X-ray dominated
regions might reveal deeply buried AGNs \citep{ran11}.
\clearpage



\clearpage

\begin{deluxetable}{ccccrc}
\tabletypesize{\small}
\tablecolumns{6}
\tablecaption{NVSS-detected QSOs, $0.2 < z < 0.45$ \label{nvsstable}}
\tablewidth{270pt}
\tablehead{ 
\multicolumn{2}{c}{ } & \colhead {Redshift} & \colhead { } & 
\colhead{~~$S$} &\colhead {$\log L_{1.4}$} \cr
\multicolumn{2}{c}{RA (J2000) DEC} & \colhead {$z$} & 
\colhead {~~$M_{\rm i}$} & \colhead {~(mJy{\rlap)}} &
  \colhead {(W~Hz$^{-1}$)}
}
\startdata
07 54 03.61 & +48 14 28.1 & 0.2755& $-$23.478 &      7.3 & 24.19 \\
08 06 44.43 & +48 41 49.2 & 0.3701& $-$23.769 &    901.7 & 26.57 \\
08 20 15.61 & +59 42 28.4 & 0.3676& $-$25.461 &      5.6 & 24.36 \\
08 33 53.88 & +42 24 01.9 & 0.2491& $-$23.903 &    248.6 & 25.62 \\
08 36 58.91 & +44 26 02.3 & 0.2544& $-$24.999 &      6.6 & 24.07 \\
08 43 10.79 & +39 53 45.1 & 0.4036& $-$23.765 &      2.6 & 24.12 \\
08 43 47.85 & +20 37 52.5 & 0.2273& $-$23.034 &     59.7 & 24.92 \\
08 49 40.01 & +09 49 21.1 & 0.3656& $-$23.425 &    587.2 & 26.37 \\
08 50 39.96 & +54 37 53.3 & 0.3673& $-$23.076 &    166.5 & 25.83 \\
08 55 16.21 & +56 16 56.8 & 0.4422& $-$23.759 &      5.3 & 24.52 \\
08 56 32.99 & +59 57 46.9 & 0.2830& $-$23.858 &    239.4 & 25.73 
\enddata
\tablecomments{Table \ref{nvsstable} is published in its entirety in the 
electronic edition of the {\it Astrophysical Journal}.  A portion is 
shown here for guidance regarding its form and content.}

\end{deluxetable}

\begin{deluxetable}{ccc}
\tabletypesize{\small}
\tablecolumns{3}
\tablecaption{NVSS ``snapshot bias'' \label{biastable}}
\tablewidth{0pt}
\tablehead{
\colhead {FLS flux density} & \colhead {Number of} & \colhead {NVSS bias} \cr
\colhead {($\mu$Jy)} & \colhead {sources} & 
\colhead {($\mu{\rm Jy~beam}^{-1}$)} 
}
\startdata
115--160 & 937 & $-17 \pm 15$ \\
160--230 & 821 & \hphantom{2}$+7 \pm 16$ \\
230--460 & 771 & $+29 \pm 17$ \\
115--460 & {\llap2}550 & $+3 \pm 9$ 
\enddata
\end{deluxetable}

\begin{deluxetable}{ccccrc}
\tabletypesize{\small}
\tablecolumns{6}
\tablecaption{NVSS-detected QSOs, $1.8 < z < 2.5$ \label{nvsshiztable}}
\tablewidth{270pt}
\tablehead{ 
\multicolumn{2}{c}{ } & \colhead {Redshift} & \colhead { } & 
\colhead{~~$S$} &\colhead {$\log L_{1.4}$} \cr
\multicolumn{2}{c}{RA (J2000) DEC} & \colhead {$z$} & 
\colhead {~~$M_{\rm i}$} & \colhead {~(mJy{\rlap)}} &
  \colhead {(W~Hz$^{-1}$)}
}
\startdata
08 06 20.47 & $+$50 41 24.4 & 2.4565 & $-$28.38 &    16.9 & 26.75 \\
08 11 41.98 & $+$51 57 10.9 & 2.1763 & $-$27.65 &     5.1 & 26.12 \\
08 21 53.82 & $+$50 31 20.4 & 2.1326 & $-$27.93 &    58.5 & 27.16 \\
08 27 54.25 & $+$33 36 04.2 & 1.8056 & $-$27.04 &    31.5 & 26.72 \\
08 33 50.61 & $+$38 39 22.8 & 2.0143 & $-$27.53 &     2.7 & 25.76 \\
08 34 00.05 & $+$43 01 48.1 & 2.3916 & $-$27.39 &    19.0 & 26.78 \\
08 37 22.41 & $+$58 25 01.8 & 2.1010 & $-$27.89 &   690.4 & 28.21 \\
08 45 06.24 & $+$42 57 18.4 & 2.0945 & $-$27.38 &   224.0 & 27.72 \\
08 45 47.19 & $+$13 28 58.1 & 1.8834 & $-$28.40 &   413.4 & 27.88 \\
08 50 51.80 & $+$15 22 15.0 & 2.0183 & $-$27.85 &   520.2 & 28.05 
\enddata
\tablecomments{Table \ref{nvsshiztable} is published in its entirety in the 
electronic edition of the {\it Astrophysical Journal}.  A portion is 
shown here for guidance regarding its form and content.}
\end{deluxetable}


\begin{thebibliography}{}
\bibitem[Balokovi\'{c} et al.(2012)]{bal12}
        Balokovi\'{c}, M., Smol\v{c}i\'{c}, V., Ivezi\'{c}, Z., et al.~2012,
        ApJ, in press (arXiv:1209.1099B)
\bibitem[Barthel(2006)]{bart06}
        Barthel, P.~D. 2006, A\&A, 458, 1007
\bibitem[Barvainis(1990)]{barv90}
        Barvainis, R. 1990, \apj, 353, 419
\bibitem[Barvainis et al.(2005)]{barv05}
        Barvainis, R., Leh\'{a}r, J., Birkinshaw, M., Falcke,
        H., \& Blundell, K.~M. 2005, \apj, 618, 108
\bibitem[Becker, White, \& Helfand(1995)]{bec95}
        Becker, R.~H., White, R.~L., \& Helfand, D.~J. 1995, \apj, 450, 559
\bibitem[Blundell \& Beasley(1998)]{blu98}
        Blundell, K.~M., \& Beasely, A.~J. 1998, MNRAS, 299, 165
\bibitem[Blundell \& Kuncic(2007)]{blu07}
        Blundell, K.~M., \& Kuncic, Z. 2007, ApJ, 668, L103
\bibitem[Bonfield et al.(2011)]{bon11}
        Bonfield, D.~G., Jarvis, M.~J., Hardcastle, M.~J.~et al.
        2011, MNRAS, 416, 13
\bibitem[Boyce et al.(1998)]{boy98}
        Boyce, P.~J., Disney, M.~J., Blades, J.~C., et al.
        1998, MNRAS, 298,121
\bibitem[Condon(1992)]{con92}
        Condon, J.~J. 1992, ARA\&A, 30, 575
\bibitem[Condon, Cotton, \& Broderick(2002)]{con02}
        Condon, J.~J., Cotton, W.~D., \& Broderick, J.~J. 2002,
        \aj, 124, 675
\bibitem[Condon et al.(1991)]{con91}
        Condon, J.~J., Huang, Z.-P., Yin, Q.~F., \& Thuan, T.~X.~T.
        1991, ApJ, 378, 65
\bibitem[Condon \& Mitchell(1984)]{con84}
        Condon, J.~J., \& Mitchell, K.~J. 1984, AJ, 89, 610
\bibitem[Condon et al.(1998)]{con98}
        Condon, J.~J., Cotton, W.~D., Greisen, E.~W., et al. 1998, 
        \aj, 115, 1693 (NVSS)
\bibitem[Condon et al.(2003)]{con03}
        Condon, J.~J., Cotton, W.~D., Yin, Q.~F., et al. 2003,
        AJ, 125, 2411
\bibitem[Cotton et al.(1980)]{cot80}
        Cotton, W.~D., Wittels, J.~J., Shapiro, I.~I. et al. 1980,
        ApJ, 238, L123
\bibitem[Crawford et al.(1970)]{cra70}
        Crawford, D.~F., Jauncey, D.~L., \& Murdoch, H. ~S. 1970,
        ApJ, 162, 405
\bibitem[Croton et al.(2006)]{cro06}
        Croton, D.~J., Springel, V., White, S.~D.~M.~et al.
        2006, MNRAS,365, 11
\bibitem[de Vries et al.(2007)]{dev07}
        de Vries, W.~H., Hodge, J.~A., Becker, R.~H., White, R.~L.,
        \& Helfand, D.~J. 2007, AJ, 134, 457
\bibitem[Dunlop et al.(2003)]{dun03}
        Dunlop, J.~S., McLure, R.~J., Kukula, M.~J.,
        et al. 2003, MNRAS, 340, 1095
\bibitem[Falcke et al.(1996)]{fal96}
        Falcke, H., Patnaik, A.~R., \& Sherwood, W. 1996, ApJ, 473, L13
\bibitem[Greenstein \& Matthews(1963)]{gre63}
        Greenstein, J.~L., \& Matthews, T.~A. 1963, Nature, 197, 1041
\bibitem[Hazard et al.(1963)]{haz63}
        Hazard, C., Mackey, M.~B., \& Shimmins, A.~J. 1963,
        Nature, 197, 1037
\bibitem[Hoyle \& Burbidge(1966)]{hoy66}
        Hoyle, F., \& Burbidge, G.~R. 1966, ApJ, 144, 534
\bibitem[Ivezi\'{c} et al.(2002)]{ive02}
        Ivezi\'{c}, Z., Menou, K., Knapp, G.~R., et al.~2002, AJ, 124, 2364
\bibitem[Kamionkowski \& Loeb(1997)]{kam97}
        Kamionkowski, M., \& Loeb, A.~1997, Phys.Rev.D, 56, 4511
\bibitem[Katgert et al.(1973)]{kat73}
        Katgert, P., Katgert-Merkelijn, J.~K., Le Poole, R.~S., 
        \& van der Laan, H. 1973, A\&A, 23, 171
\bibitem[Kellermann \& Pauliny-Toth(1966)]{kel66}
        Kellermann, K.~I., \& Pauliny-Toth, I.~I.~K. 1966, 
        Nature, 212, 781
\bibitem[Kellermann et al.(1989)]{kel89}
        Kellermann, K.~I., Sramek, R., Schmidt, M., Shaffer, D.~B., 
        \& Green, R. 1989, \aj, 98, 1195
\bibitem[Kellermann et al.(1994)]{kel94}
        Kellermann, K.~I., Sramek, R.~A., Schmidt, M., Green, R.~F.,
        \& Shaffer, D.~B. 1994, \aj, 108, 1163
\bibitem[Kimball et al.(2011)]{kim11}
        Kimball, A.~E., Kellermann, K.~I., Condon, J.~J., Ivezi\'{c},
        Z., \& Perley, R.~A. 2011, \apj, 739, L29
\bibitem[Mahony et al.(2012)]{mah12}
        Mahony, E., Sadler, E.~M., Croom, S.~M.~et al.~2012,
        ApJ, 754, 12
\bibitem[Matthews \& Sandage(1963)]{mat63}
        Matthews, T.~A., \& Sandage, A.~R. 1963, ApJ, 138, 30
\bibitem[Miller et al.(1990)]{mil90}
        Miller, L., Peacock, J.~A., \& Mead, A.~R.~G. 1990,
        MNRAS, 24, 207
\bibitem[Miller et al.(1993)]{mil93}
        Miller, P., Rawlings, S., \& Saunders, R. 1993, MNRAS,
        263, 425
\bibitem[Mitchell \& Condon(1985)]{mit85}
        Mitchell, K.~J., \& Condon, J.~J. 1985, AJ, 90, 1957
\bibitem[Mori\'{c} et al.(2010)]{mor10}
        Mori\'{c}, L., Smol\v{c}i\'{c}, V., Kimball, A., et al. 2010, 
        ApJ, 724, 779
\bibitem[Mundell, Ferruit, \& Pedlar(2001)]{mun01}
        Mundell, C.~G., Ferruit, P., \& Pedlar, A. 2001, ApJ, 560, 168
\bibitem[Peacock et al.(1986)]{pea86}
        Peacock, J.~A., Miller, L., \& Longair, M.~S. 1986, 
        MNRAS, 218, 265
\bibitem[Rangwala et al.(2011)]{ran11}
        Rangwala, N., Maloney, P.~R., Glenn, J., et al. 2011,
        ApJ, 743:94
\bibitem[Rovilos et al.(2012)]{rov12}
        Rovilos, E., Comastri, A., Gilli, R.~et al.~2012,
        A\&A, in press (arXiv:1207.7129)
\bibitem[Sandage(1965)]{san65}
        Sandage, A. 1965, \apj, 141, 1560
\bibitem[Scheuer \& Readhead(1979)]{sch79}
        Scheuer, P.~A.~G., \& Readhead, A.~C.~S. 1979,
        Nature, 277, 182
\bibitem[Schmidt(1963)]{sch63}
        Schmidt, M. 1963, Nature, 197, 1040
\bibitem[Schmidt(1970)]{sch70}
        Schmidt, M. 1970, \apj, 162, 371
\bibitem[Schmidt \& Green(1983)]{sch83}
        Schmidt, M., \& Green, R.~F. 1983, \apj, 269, 352
\bibitem[Schneider et al.(2010)]{sch10}
        Schneider, D.~P., Richards, G.~T., Hall, P.~B.~et al. 
        2010, \aj, 139, 2360
\bibitem[Sopp \& Alexander(1991)]{sop91}
        Sopp, H.~M., \& Alexander, P. 1991, MNRAS, 251, 14P
\bibitem[Steenbrugge et al.(2011)]{ste11}
        Steenbrugge, K.~C., Jolley, E.~J.~D., Kuncic, Z., \&
        Blundell, K.~M. 2011, MNRAS, 413, 1735
\bibitem[Strittmatter et al.(1980)]{str80}
        Strittmatter, P.~A., Hill, P., Pauliny-Toth, I.~I.~K.,
        Steppe, H., \& Witzel, A. 1980, A\&A, 88, L12
\bibitem[Ulvestad et al.(2005)]{ulv05}
        Ulvestad, J.~S., Antonucci, R.~R.~J., \& Barvainis, R.
        2005, ApJ, 621, 123 
\bibitem[Walsh et al.(1989)]{wal89}
        Walsh, D.~E.~P., Knapp, G.~R., Wrobel, J.~M., \& 
        Kim, D.-W. 1989, \apj, 337, 209
\bibitem[White et al.(1997)]{whi97}
        White, R.~L., Becker, R.~H., Helfand, D.~J., \&
        Gregg, M.~D. 1997, ApJ, 475, 479
\bibitem[White et al.(2007)]{whi07}
        White, R.~L., Helfand, D.~J., Becker, R.~H., Glikman, E.,
        \& de Vries, W. 2007, ApJ, 654, 99
\end{thebibliography}
\end{document}